\documentclass[lettersize,journal,a4paper]{IEEEtran}

\usepackage{amsmath}
\usepackage{amsthm}

\theoremstyle{plain}

\newtheorem{remark}{Remark}[section]

\newtheorem{proposition}{Proposition}[section]

\usepackage{color}
\usepackage{graphicx} 
\usepackage{bm}
\usepackage{multicol}
\usepackage{setspace}
\usepackage{subfigure}
\usepackage{epstopdf}
\usepackage{fancyhdr} 
\usepackage{indentfirst}
\usepackage{enumerate}
\usepackage{amssymb}
\usepackage{stfloats}
\usepackage{bm}
\usepackage{threeparttable}
\usepackage{CJK}
\usepackage[justification=centering]{caption}
\usepackage{makecell}
\usepackage{caption}
\usepackage{cases}
\usepackage{algorithm}
\usepackage{algpseudocode}
\usepackage{graphics}
\usepackage{subfig}
\usepackage{epsfig}
\usepackage{cite}

\begin{document}
	\title{Visual Fidelity Index for Generative Semantic Communications with Critical Information Embedding }
	\author{
		{Jianhao~Huang, Qunsong Zeng, \IEEEmembership{Member,~IEEE}, and Kaibin Huang, ~\IEEEmembership{Fellow,~IEEE} }
		\thanks{
			
			{J. Huang, Q. Zeng, and K. Huang are with the Department of Electrical and Electronic Engineering, The University of Hong Kong, Hong Kong.  Emails: \{jianhaoh,  huangkb\}@hku.hk, qszeng@eee.hku.hk. Corresponding author: K. Huang. }

		}
	}
	
	\maketitle
	\thispagestyle{empty}

\begin{abstract}
Generative semantic communication (Gen-SemCom) with large artificial intelligence (AI) model promises a transformative paradigm for 6G networks, which reduces communication costs by transmitting low-dimensional prompts rather than raw data. However, purely prompt-driven generation loses fine-grained visual details. Additionally, there is a lack of  systematic metrics  to evaluate the performance of  Gen-SemCom systems.  To address these issues, we develop a hybrid Gen-SemCom system with a critical information embedding (CIE) framework, where both text prompts and semantically critical features are extracted for transmissions.   First,  a novel approach of semantic filtering  is proposed to select and transmit the semantically critical features of images relevant to semantic label.  By integrating the text  prompt and  critical features, the receiver reconstructs high-fidelity images using a diffusion-based generative model. 
Next, we propose the generative visual information fidelity (GVIF) metric to evaluate the visual quality of the generated image. By characterizing the statistical models of image features,  the GVIF metric quantifies the mutual information  between the distorted features and their original counterparts. 
By maximizing the GVIF metric, we  design a channel-adaptive Gen-SemCom system that adaptively control the volume of features and compression rate according to the channel state. Experimental  results validate the GVIF metric’s sensitivity to visual fidelity, correlating  with both the peak signal-to-noise ratio (PSNR) and critical information volume. In addition, the optimized system achieves superior performance over benchmarking schemes in terms of  higher PSNR and lower Fréchet Inception Distance (FID) scores. 

\end{abstract}


\begin{figure}[t]
	\centering
	\includegraphics[width=1.0\linewidth]{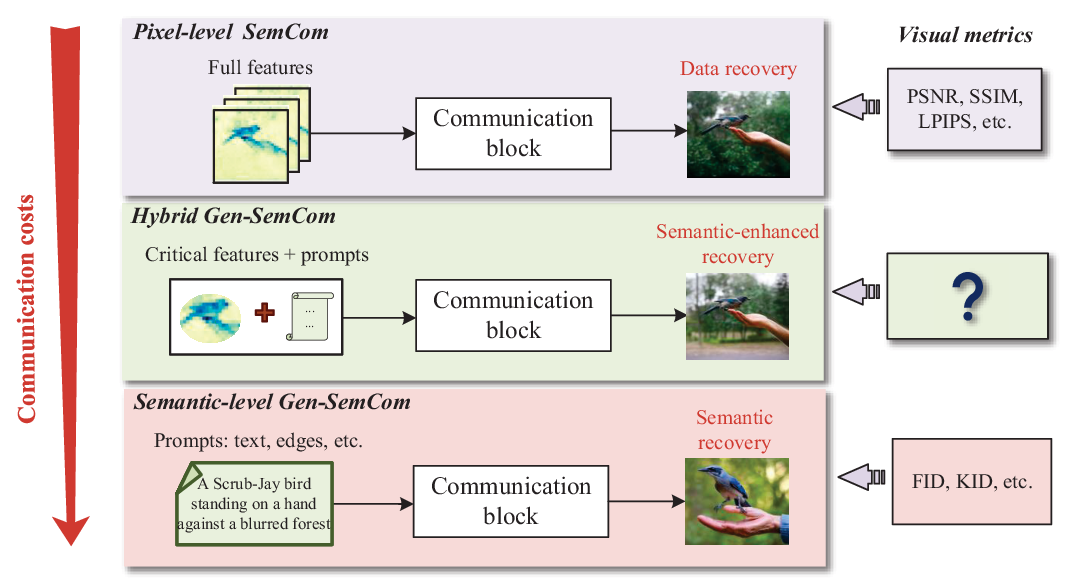}
	\captionsetup{justification=justified}
	\caption{Comparisons of multi-level SemCom systems with different visual metrics. }
	\label{intro}
\end{figure}

\section{Introduction}

Semantic communication (SemCom) is envisioned to be a distinctive paradigm for  sixth-generation (6G) wireless networks, which redefines  transmission protocols by emphasizing the conveyance of semantic meanings of source data \cite{gunduz2022beyond,zhang2022toward,10287247}. Unlike traditional systems that focus on raw-data transmission, SemCom leverages advanced analysis of source data to identify contextually critical features, thereby  reducing communication overhead. This involves data analytics across syntactic and semantic levels to ensure only task-relevant information, such as objects in an image or intent in speech, is transmitted. For applications like autonomous vehicles, industrial Internet-of-Things (IoT), and extended reality, SemCom can achieve more efficient spectrum utilization and lower latency, aligning with the 6G vision of ultra-fast, intelligent connectivity \cite{saad2019vision,zhu2020toward,qu2025partialloading,10529950,10415235}. 


Current research in SemCom  revolves around syntactic-level (or pixel-level) transmission, where variational autoencoders (VAEs) compress  data by optimizing a trade-off between recovery distortion and compression rate \cite{Balle2017,Balle2018,liu2023learned,10175391}. 
Most research focuses on exploring advanced VAE architectures, e.g., convolutional neural network (CNN) and Transformer,  to improve the  rate-distortion trade-offs \cite{Balle2018,liu2023learned,10175391, dai2022nonlinear}. On one hand, hierarchical VAEs utilize multi-scale CNNs to compress images into the latent feature domain  \cite{Balle2018}. On the other hand, the transformer-augmented VAEs exploit self-attention mechanisms to capture long-range dependencies, achieving near-optimal rate-distortion performance \cite{liu2023learned,li2023fundamental}. Considering  transmission errors, researchers have integrated VAE-based architectures into joint source-channel coding frameworks, yielding a Deep Joint Source-Channel Coding (DJSCC) scheme that balances  the end-to-end distortion and transmission rate \cite{ bourtsoulatze2019deep,10845799}. For scenarios involving extreme channel fading,  the VAE decoder can be combined with a Generative Adversarial Network (GAN) to enhance the reconstructed visual quality \cite{erdemir2023generative}. 
These approaches, however, remain constrained by their reliance on pixel-level metrics like peak signal-to-noise ratio (PSNR) and structural similarity index (SSIM), which emphasize syntactic accuracy over semantic relevance.   Although the metric of learned perceptual image patch similarity (LPIPS)  was proposed to better capture structural similarity in images, it fails to account for semantic content (e.g.  objects) \cite{zhang2018unreasonable}. For instance, disregarding  redundant background pixels in a video call may not change its semantic content but degrade the pixel-level metrics. The limitations of traditional approaches necessitate a paradigm shift toward semantic-level processing. 

Generative SemCom (Gen-SemCom) has emerged as a transformative approach to address these limitations by leveraging generative AI (GenAI) to reconstruct data from given semantic meanings \cite{10531769, zhang2025semantics, grassucci2023generative,  jiang2024large, jiang2025m4sc,10726905}. Unlike pixel-level transmissions, semantic-level Gen-SemCom encodes data  into low-dimensional  prompts (such as text descriptions \cite{10531769,zhang2025semantics,lin2025hsplitlora}, semantic layouts \cite{grassucci2023generative}, or edge maps \cite{zhang2025semantics}) for  transmissions, and uses a generative model at the receiver to generate semantically accurate outputs \cite{rombach2022high, zhao2023survey}.  As the size of prompts is much smaller than image's full features, this approach can significantly reduce transmission costs.	In particular, a diffusion-based SemCom framework has been proposed to progressively generate high-fidelity images from transmitted semantic layout through iterative denoising steps  \cite{grassucci2023generative}. The approach has been demonstrated to  achieve 92\% reduction in bit number while maintaining a comparable semantic accuracy as the pixel-level methods. However, purely prompt-driven generation risks losing fine-grained visual fidelity (see Fig. \ref{intro}), as subtle textures or domain-specific details (e.g., medical imaging anomalies) may be oversimplified in the generative process.  In addition, conventional metrics, e.g. PSNR and LPIPS, are inadequate    for quantifying the quality of generative images. Evaluating such generative systems usually adopts visual metrics like  Kernel Inception Distance (KID) and Fréchet Inception Distance (FID), which assess statistical similarity between generated and real data distributions \cite{heusel2017gans}. However, these metrics are computationally prohibitive and unsuitable for per-sample quality assessment. 

To address these issues, hybrid Gen-SemCom system has manifested  as an in-between framework, unifying the efficiency of semantic-level prompts with the reliability of  syntactic-level transmission \cite{thorsager2024generative, du2024generative,tariq2023segment}. 
By strategically transmitting  prompts alongside with  critical pixels of data, the hybrid Gen-SemCom system balances spectrum efficiency and reconstruction fidelity. The key component of  this framework is \emph{critical information embedding} (CIE), a process that selectively transmits semantically critical pixel information  alongside extracted prompts.  This embedded design is particularly vital for video and medical applications, where it is important to retain critical pixels for anchoring the generation process to ground-truth details. However, existing work  transmit either all image information \cite{du2024generative}, which is inefficient,  or randomly selected pixels \cite{thorsager2024generative}. There lacks an effective and controllable approach to extract and transmit semantically critical information.  Although segmentation-based methods can  identify object pixels, they are computationally expensive \cite{kirillov2023segment,tariq2023segment}. Furthermore, these methods lack the flexibility to  adjust the volume of transmitted  information based on  channel conditions.
More importantly, there exists no suitable metric to  evaluate the hybrid Gen-SemCom systems, as conventional ones like FID or PSNR fail to relate the visual fidelity with semantically critical information.


In this paper, we focus on the hybrid Gen-SemCom system with the CIE  for wireless image transmission. In this system, the large language model and the VAE approach are combined together for prompt extraction and projection of images into a low-dimensional latent feature space for efficient CIE processing.
The main contributions and findings of this paper are summarized as follows. 

\begin{itemize}
	\item \textbf{Semantic filtering}:  First, we propose the application of semantic filtering  to enable the CIE process by selecting semantically critical features for transmission.  First, our approach leverages the spatial consistency of CNNs to model feature importance. To this end,  class activation mapping (CAM) \cite{zhou2016learning} is adopted to generate spatial importance matrices that highlight regions relevant to semantic labels. Building on this, we introduce a threshold-based filtering mechanism that selectively discards features with importance scores below a tunable threshold 
$\alpha$ to reduce transmission overhead. The threshold serves to balance efficiency and fidelity: a higher value reduces communication overhead, while a lower one preserves richer original visual information.
	
    \item \textbf{Visual fidelity index}: 
    Next, we propose the generative visual information fidelity (GVIF) metric to evaluate the visual quality of the generated images. Different from  pixel-level metrics, the GVIF metric focuses on the amount of information  persevered in the features of generated image compared to the original counterparts, which is meant for human vision system (HVS)~\cite{sheikh2006image}. 
	To design the GVIF, we first model the distribution of image features using the Gaussian Scale Mixture (GSM) model and also derive  distortion models introduced by various system operations, i.e, VAE-based compression, semantic filtering, and generation. Then, the HVS is modeled as a ``Gaussian channel" by adding Gaussian noises into features, which imposes limits on the information flows into human brain \cite{sheikh2006image}.   Finally, the GVIF calculates the normalized mutual information between the distorted features and their original counterparts.  The proposed GVIF metric is shown to be related to the  source rate and number of transmitted features, both of which contribute to the total transmission cost of the Gen-SemCom system. 
	\item \textbf{Application to channel-adaptive Gen-SemCom}:   Finally, we optimize the performance of the hybrid Gen-SemCom system by maximizing the proposed GVIF metric according to the channel state. For a given receive signal-to-noise ratio (SNR), the VAE-based source coder and  the filtering threshold $\alpha$ are jointly optimized to maximize the average GVIF under a latency constraint. To efficiently solve this problem, we propose a two-step algorithm: First, fix the VAE-based source coder and optimize the threshold $\alpha$ by applying zero-order gradient descent method; Then, select the best source coder from a list of pre-trained models with different rate-distortion pairs.  
	\item \textbf{Experimental results}: The experimental results reveal that the proposed GVIF metric has strong capability to quantify the visual fidelity of the generated images. In particular, we observe that GVIF exhibits positive correlation with the PSNR and  volume of  critical features, validating its sensitivity to visual fidelity. By optimizing the hybrid Gen-SemCom framework using GVIF, the system outperforms JPEG2000 and VAE-based benchmarkings, achieving higher PSNR in critical regions. Furthermore,  the GVIF-driven framework reduces semantic ambiguity, yielding notably lower FID scores than prompt-to-image-only generation. 
\end{itemize}

The remainder of this paper is organized as follows. The system model of the hybrid Gen-SemCom is introduced in Section II. The proposed semantic filtering is presented in Section III. The GVIF metric is proposed in Section IV. Experimental results are presented in Section V, followed by concluding remarks in Section VI.

Notations:  We utilize lowercase and uppercase letters, e.g., $x$ and $M$, to denote  scalars, and use boldface  letters, e.g., $\bm{x}$ and $\bm{X}$, to denote vectors, matrices, and tensors. $\mathbb{Z}$,  $\mathbb{R}$, and  $\mathbb{C}$, denote the sets of all integer, real, and complex values, respectively.
$\|\bm{x}\|$ denotes the $2$-norm  of  $\bm{x}$. $p(\bm{x})$ denotes the probability density function (PDF) of the continuous random variable $\bm{x}$.  $P(\bm{y})$ denotes the probability mass function (PMF) of the discrete random variable $\bm{y}$. 
$\mathbb{E}_{\bm{y}}(\cdot)$  denotes the expectation over  random variable ${\bm{y}}$.   $O(\cdot)$ denotes the big O  notation.  $\text{log}_2(\cdot)$ is the logarithm function with base $2$. $[N]$ denotes the integer  set $\{1,2,\cdots,N\}$.

\section{Hybrid Gen-SemCom System}
\begin{figure*}[t]
	\normalsize
	\setlength{\abovecaptionskip}{+0.3cm}
	\setlength{\belowcaptionskip}{-0.1cm}
	\centering
	\includegraphics[width=0.87\linewidth]{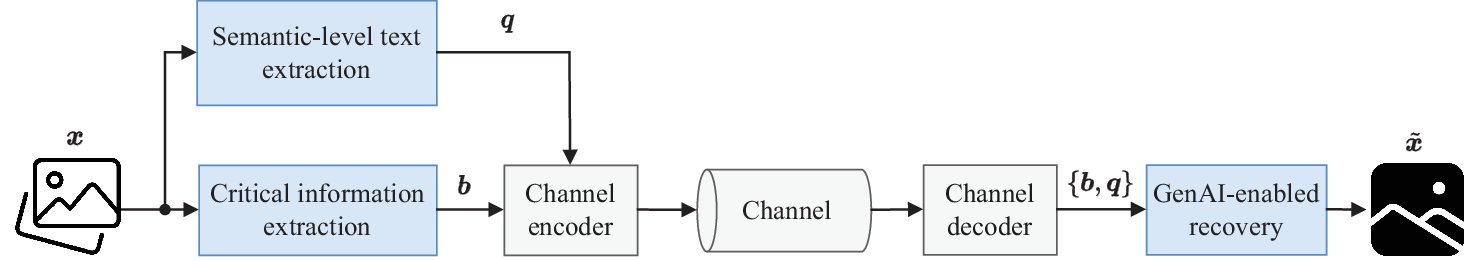}
	\captionsetup{justification=justified}
	\caption{The hybrid Gen-SemCom system with the CIE process. }
	\label{Neural_Arch1}
\end{figure*}

Consider a hybrid Gen-SemCom system for wireless image transmissions, as shown Fig. \ref{Neural_Arch1}, where only semantic-critical information is extracted from raw data to guide high-fidelity image generation at the receiver. Central to this system is the CIE process. The CIE process, along with other system operations and models, is described as follows.

\subsection{Critical Information Embedding}
   The CIE process  encodes and transmits two  components: (1) semantic-level  prompts (e.g., text descriptions and layouts) and (2) critical pixel information.  By embedding these two components, the receiver utilizes generative AI to output semantically accurate images. The details are provided as follows.

 \begin{figure*}[t]
	\normalsize
	\setlength{\abovecaptionskip}{+0.3cm}
	\setlength{\belowcaptionskip}{-0.1cm}
	\centering
	\includegraphics[width=1.\linewidth]{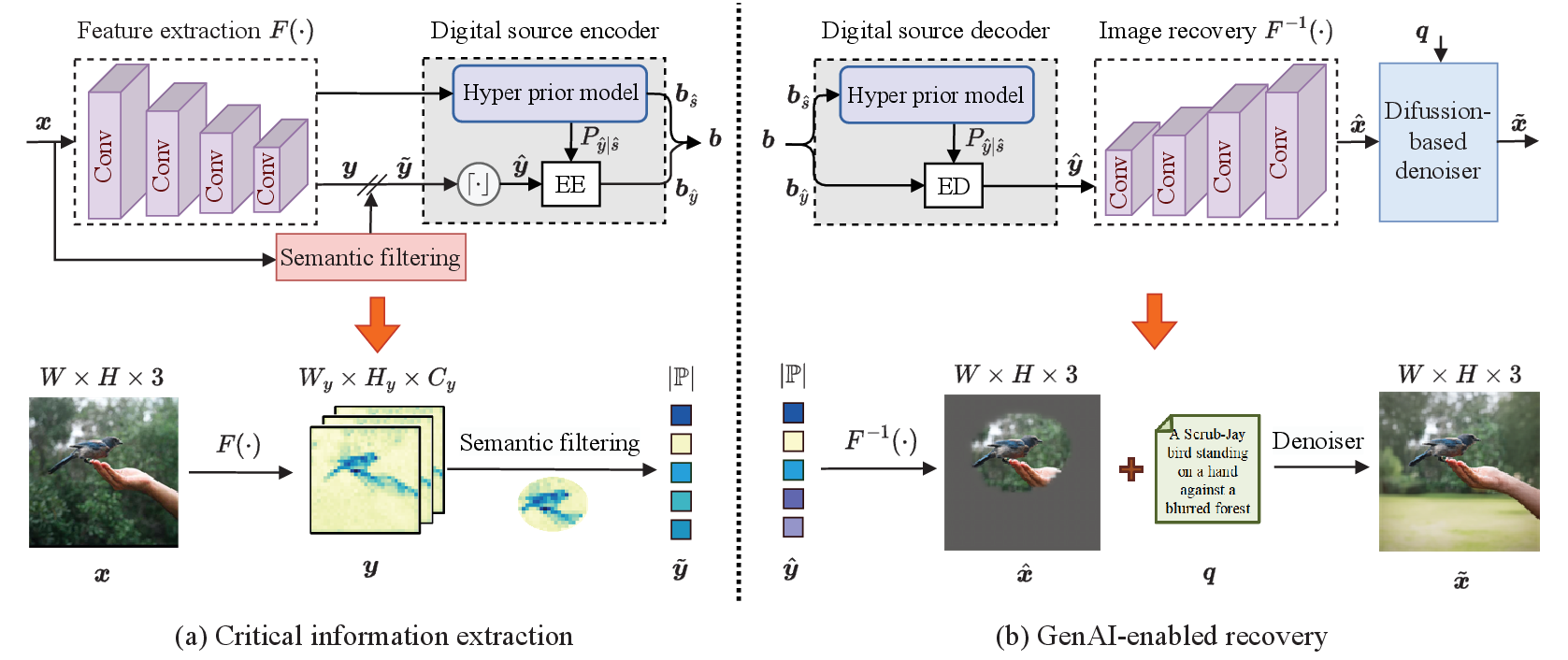}
	\captionsetup{justification=justified}
	\caption{ The system operations of the CIE process. Conv, $\left\lceil \cdot \right\rfloor$, EE, and ED represent the convolutional layer,  scalar quantization, entropy encoder, and entropy decoder, respectively. }
	\label{Neural_Arch2}
\end{figure*}

\subsubsection{Semantic-level  prompts}  In this paper, we consider the widely-used text descriptions as the prompt. Text descriptions  encapsulate the coarse but  high-level semantic information behind images. Modern large AI based language models like GPT-4 \cite{chatgpt2023,qu2025mobile}, BLIP-2 \cite{li2023blip2}, can automatically generate detailed captions that describe image content, spatial relationships, and even abstract concepts present in visual scenes. Such text-based representations, while not capturing every fine-grained visual detail, require minimal bandwidth for transmission—typically just a few hundred bytes compared to the hundreds of kilobytes or megabytes needed for the original images.   With the image sample $\bm{x}$, the text extraction process can be parameterized by 
\begin{align}
	\bm{q}=\mathcal{H}(\bm{x};\bm{\Gamma}),
\end{align}
where $\bm{\Gamma}$ is the parameter set of NNs.  

\subsubsection{Critical pixel information} 
To improve the quality of image generation, we consider not only using text descriptions but also embedding partial yet critical pixels of images. These pixels provide critical visual details that improve the accuracy and realism of the generated images. However, transmitting these pixels comes with a cost, which is not neglectable compared to the cost of transmitting text descriptions. Hence, additional compression and coding is prerequisite to process the images before transmission. Here, we denote the coding process as 
\begin{align}
	\bm{b}=\mathcal{F}(\bm{x};\bm{\Phi}),
\end{align}
where $\bm{b}$ is the coded bits of critical information and $\bm{\Phi}$ is the set of NN parameters.
Since text prompts  provide only semantic-level descriptions, the visual  fidelity of generated images is mainly constrained by the  transmission of critical pixels. This  poses a central challenge in designing effective encoding and transmission framework for the critical pixel  information.


\subsection{System Operations and Models}
In this subsection, we introduce an effective  framework design for the hybrid Gen-SemCom with CIE.
In particular, we utilize the well-trained BLIP model \cite{li2023blip2} to obtain the text descriptions, $\bm{q}$, whose transmission cost is omitted.  The encoding  and transmissions of critical pixels are illustrated as follows.

\subsubsection{Critical information extraction}

Consider an  image sample $\bm{x}\in \mathbb{R}^{W \times H \times 3}$, taking values  from the range $[0,255]$ with $W$ and $H$ being the width and height, respectively.  For critical pixels,  we adopt the VAE-based scheme to achieve sufficient encoding, as shown in Fig. \ref{Neural_Arch2}. First, the image sample, $\bm{x}$, is input into a CNN-based encoder function, denoted by $F(\cdot)$, to extract its continuous feature $\bm{y} \in \mathbb{R}^{W_y\times H_y \times C_y}$ with $W_y$, $H_y$,  and $C_y$ being the width, height, and number of feature maps of feature, respectively. In another word, 
\begin{align}
	\bm{y}=F(\bm{x};\Phi_E),
\end{align}
where $\Phi_E$ denotes the  parameter set of the CNN-based function. Let $y_{ijc}$ denote the $(i,j,c)$-th element of feature $\bm{y}$.  Then, $\bm{y}$ is fed into a \emph{semantic filtering} process to select the critical feature elements corresponding to the semantic information, which will be introduced in Section III. The filtered feature element follows  $\tilde{y}_{ijc}=y_{ijc}$, $\forall (i,j,c)\in \mathbb{P}$, with $\mathbb{P}$ being the selected index set. Next, the continuous  element $\tilde{y}_{ijc}$ is quantized  as a discrete-valued element $\hat{y}_{ijc} \in \mathbb{Z}$ using a scalar uniform quantizer with unit step size  \cite{Balle2018}, which  operates as $\hat{y}_{ijc}=\left\lceil \tilde{y}_{ijc} \right\rfloor$.  Then, the integer elements $\{\hat{y}_{ijc}\}$ are encoded into a bit stream $\bm{b}_{\hat{y}}$ by an entropy encoder, e.g., arithmetic encoder \cite{witten1987arithmetic},  according to its PMF  $P_{\hat{\bm{y}}}(\hat{\bm{y}})$.   


To accurately calculate the PMF, we adopt the \emph{hyper prior model} \cite{Balle2018},
which assumes the Gaussian distributed nature of these feature elements with their variances estimated by NNs. First,  the continuous-valued feature $\bm{y}$ is fed into a function $L_1(\bm{y};{\Omega_1})$ with parameter $\Omega_1$ to extract the latent feature, $\bm{s} \in \mathbb{R}^{W_s\times H_s \times C_s}$, which is quantized as  $\hat{\bm{s}}\in \mathbb{Z}^{W_s\times H_s \times C_s}$, i.e., $\hat{\bm{s}}=\left\lceil \bm{s} \right\rfloor$. The dimension of $\hat{\bm{s}}$ is much smaller than that of $\bm{y}$. The filtered feature elements, $\{\tilde{y}_{ijc}\}$, conditioned on  $\hat{\bm{s}}$ can be modeled as the independent while not identically distributed  (i.n.i.d.) Gaussian random variable  with zero mean and  variance $\theta^2_{ijc}$. In other words, 
\begin{align} \label{norm:1}
	p_{\tilde{y}_{ijc}|\hat{\bm{s}}}(\tilde{y}_{ijc}|\hat{\bm{s}})=\mathcal{N}(\tilde{y}_{ijc};0,\theta^2_{ijc}), (i,j,c) \in \mathbb{P},
\end{align}
where $\mathcal{N}(a;u,\theta^2)$ denotes the  PDF of a Gaussian distribution with mean $u$ and variance $\theta^2$, evaluated at the point $a$.  
$\{\theta_{ijc}\}$ are estimated by applying a transform function $L_{2}(\cdot;\Omega_2)$ with parameter $\Omega_2$ to $\hat{\bm{s}}$, i.e., $\bm{\theta}=L_{2}(\hat{\bm{s}};\Omega_2)$, where  $\theta_{ijc}$ is the $(i,j,c)$-th element of tensor  $\bm{\bm{\theta}} \in \mathbb{R}^{W_y\times H_y \times C_y}$, respectively. The conditional i.n.i.d. distribution of $\hat{y}_{ijc}$ given $\hat{\bm{s}}$ can be expressed as 
\begin{align} \label{den:y}
	P_{\hat{y}_{ijc}|\hat{\bm{s}}}(\hat{y}_{ijc}=k|\hat{\bm{s}})=\int_{k-0.5}^{k+0.5}\mathcal{N}(\tilde{y}_{ijc};0,\theta_{ijc}^2)d\tilde{y}_{ijc}, 
\end{align}
with $k \in \mathbb{Z}$.  Then, the conditional PMFs $\{P_{\hat{y}_{ijc}|\hat{\bm{s}}}(\hat{y}_{ijc}=k|\hat{\bm{s}})\}$ are input into the entropy encoder to encode the quantized feature $\hat{\bm{y}}$ into bits $\bm{b}_{\hat{y}} \in \{0,1\}^{B_{\hat{y}}}$.
The latent feature, $\hat{\bm{s}}$, is encoded into bits $\bm{b}_{\hat{s}} \in \{0,1\}^{B_{\hat{s}}}$ by the entropy encoder according to its PMF, $P_{\hat{\bm{s}}}(\hat{\bm{s}})$, which is  computed using  the non-parametric fully-factorized density model \cite{Balle2017}.  We have $B_{\hat{y}} \approx -\log_{2}P_{\hat{\bm{y}}|\hat{\bm{s}}}(\hat{\bm{y}}|\hat{\bm{s}})$ and $B_{\hat{s}} \approx -\log_2 P_{\hat{\bm{s}}}(\hat{\bm{s}})$.  
The total encoding bits for the image sample $\bm{x}$ are collected as $\bm{b}=\bm{b}_{\hat{y}} \cup \bm{b}_{\hat{s}} $ of dimension $B=B_{\hat{y}}+B_{\hat{s}}$. 
In a nutshell,  
the source rate for encoding critical information is calculated as 
\begin{align} \label{source-rate}
	\mathcal{R}=\mathbb{E}_{\bm{x}} \{-\log_2 P_{\hat{\bm{s}}}(\hat{\bm{s}}) -\log_{2}P_{\hat{\bm{y}}|\hat{\bm{s}}}(\hat{\bm{y}}|\hat{\bm{s}})\}.
\end{align}

\subsubsection{Transmission model}
After source encoding, the  bits $\bm{b}$ are encoded into a sequence of complex symbols, denoted by $\bm{g}\in \mathbb{C}^{L_x}$, for transmission. The average power for the transmitted symbols  follows $\mathbb{E}(\|\bm{g}\|^2)/L_{x}=1$. For a slow fading channel, the received signals, denoted by $\bm{r}=[r_1,r_2,\cdots,r_{L_{x}}]^T$, can be modeled as
\begin{align}
	\bm{r}=h\bm{g}+\bm{o},
\end{align}
where $h\in \mathbb{C}$ denotes the channel coefficient that remains constant during $L_x$ symbol time, and $\bm{o}=[o_1,o_2,\cdots,o_{L_{x}}]^T$ denotes the  additive white Gaussian noise (AWGN) with each element following $o_i \sim \mathcal{CN}(0,\sigma^2)$.  According to Shannon's  theorem \cite{Tse2005}, the maximal achievable  rate (in bits per second)  is given by $C=B\log_2(1+\frac{|h|^2}{\sigma^2})$ with $B$ being the bandwidth. Considering the error-free  transmission of rate $C$, the average latency (in seconds), denoted by $T_x$, is calculated by
\begin{align}
	T_x=\frac{\mathcal{R}}{B\log_2(1+\frac{|h|^2}{\sigma^2})}. 
\end{align} 

\subsubsection{GenAI-enabled recovery}
The receiver aims to recover a high-fidelity image based on the transmitted bits $\bm{b}$ and the text prompt $\bm{q}$.  As shown in Fig. \ref{Neural_Arch2}(b), the received bits $\bm{b}$ are separated into two parts: the feature bits $\bm{b}_{\hat{y}}$ and the side information bits $\bm{b}_{\hat{s}}$.  First,  $\bm{b}_{\hat{s}}$ is fed into the entropy decoder to decode the side information, $\hat{\bm{s}}$, according to the shared PMF $P_{\hat{\bm{s}}}(\hat{\bm{s}})$. Then, $\hat{\bm{s}}$ is fed into the function $L_2(\cdot;\Omega_2)$ to compute the PMF, $P_{\hat{\bm{y}}|\hat{\bm{s}}}(\hat{\bm{y}}|\hat{\bm{s}})$, given in \eqref{den:y}. With $P_{\hat{\bm{y}}|\hat{\bm{s}}}(\hat{\bm{y}}|\hat{\bm{s}})$ and  $\bm{b}_{\hat{y}}$,  the feature vector, $\hat{\bm{y}}$, is decoded by utilizing the entropy decoder.  Next,   feature $\hat{\bm{y}}$ is fed into a CNN-based decoder function, denoted by $F^{-1}(\cdot)$, to output  image $\hat{\bm{x}}$.  In other words,
\begin{align}
	\hat{\bm{x}}=F^{-1}(\hat{\bm{y}};\Phi_D),
\end{align}
where $\Phi_D$ denotes the parameter set of the decoder function.  Finally, the noisy image $\hat{\bm{x}}$ and the prompt $\bm{q}$ are fed into a Diffusion-based denoiser to generate image $\tilde{\bm{x}}$, which is detailed as follows. 

\subsubsection{Diffusion-based denoiser}
In this paper, we adopt the pretrained \emph{Diffusion Model}~\cite{lugmayr2022repaint,rombach2022high} to generate high-quality images $\tilde{\bm{x}}$ from noisy inputs $\hat{\bm{x}}$. First, we generate  a binary mask $\bar{\bm{m}}\in \{0,1\}^{W\times H \times 3}$ with value one indicating the positions of critical pixels. The positions of critical pixels will be transmitted to the receiver as  side information, which will be detailed in Section III-C. Based on the mask $\bar{\bm{m}}$, the  generation process can be formulated as a \emph{conditional inpainting} ~\cite{lugmayr2022repaint,rombach2022high}  where critical  pixels in $\hat{\bm{x}}$ and text prompt $\bm{q}$ jointly guide the generation of missing pixels.  We employ the reverse diffusion process that progressively transforms Gaussian noise $\bm{x}_T\sim \mathcal{N}(\bm{0},\bm{I})$ into image through $T$ iterative refinement steps.  At each timestep $t$, a UNet model \cite{rombach2022high} $U_{\theta,t}$ is applied to predict the noise component conditioned on both partial pixels and semantic prompts:
\begin{equation}
	\bm{\epsilon}_{\theta,t} = U_{\theta,t}\left(\bm{x}_t, t, \bar{\bm{m}}\odot \hat{\bm{x}}, \bm{q}\right),
\end{equation}
where $\bm{x}_t$ is the latent state at step $t$, and $\bar{\bm{m}} \odot \hat{\bm{x}}$ denotes the element-wise multiplication of the tensors $\bar{\bm{m}} $ and $\hat{\bm{x}}$. The reverse transition follows:

\begin{equation}
	\bm{x}_{t-1} = \frac{1}{\sqrt{\alpha_t}} \left( \bm{x}_t - \frac{\beta_t}{\sqrt{1 - \bar{\alpha}_t}} \bm{\epsilon}_{\theta,t}\right)+\sigma_t^2 \bm{z},
\end{equation}
where $\alpha_t$, $\beta_t$, $\sigma_t$ are  scheduling parameters, $\bar{\alpha}_t=\prod_{s=1}^{T}(1-\beta_{s})$, and $\bm{z} \sim \mathcal{N}(0,\bm{I})$ is additive noise. After $T$ iterations, we get $\tilde{\bm{x}}=\bar{\bm{m}} \odot \hat{\bm{x}}+(1-\bar{\bm{m}})\odot \bm{x}_0$. The reverse process constitutes a learned mapping  from the Gaussian noise to real image distribution. Hence, the generated image approximately follows the conditional distribution, i.e., $ \tilde{\bm{x}} \sim p(\bm{x}|\bar{\bm{m}}\odot \hat{\bm{x}}, \bm{q})$.

\subsection{Training  of Source Coder}



This subsection introduces the training details of the VAE-based source coder.  
Let $\bm{\Phi} = \{\Phi_E, \Phi_D, \Omega_1, \Omega_2\}$ be the complete set of neural network parameters for source encoding and decoding. The training of $\bm{\Phi}$ is independent of semantic filtering and generation processes, and follows the principles of rate-distortion theory. As formalized in rate-distortion theory \cite{gallager1968information}, the objective is to determine the minimum bitrate required to reconstruct original images with minimal distortion. To this end, we define the per-pixel distortion metric as the mean squared error (MSE):
\begin{align}
	\mathcal{D}(\bm{\Phi}) = \mathbb{E}_{\bm{x}}\left\{\frac{1}{3 \times W \times H} \|\bm{x} - \hat{\bm{x}}\|^2 \right\}.
\end{align}
The discrete quantization process is approximated as adding uniform noise with zero mean and radius $1$ to the feature elements. During training, the semantic filtering process is omitted, so $\tilde{\bm{y}} = \bm{y}$. The neural network parameters $\bm{\Phi}$ are then trained to optimize the rate-distortion trade-off:
\begin{align} \label{rate-distortion}
	\bm{\Phi} = \arg \min_{\bm{\Phi}} \mathcal{R}(\bm{\Phi}) + \lambda \mathcal{D}(\bm{\Phi}),
\end{align}
where $\lambda > 0$ is a balancing parameter. By adjusting $\lambda$, multiple source coding models can be obtained with different rate-distortion pairs.

\section{ Semantic  Filtering}

In this section, we propose the approach of  semantic  filtering  to select semantically non-critical feature elements to prune for saving transmission resources.  First, we develop a  framework to quantify feature importance with respect to (w.r.t.) the image's semantic label. Building on the feature importance, we present a feature filtering strategy that  prunes non-critical features.

\subsection{Semantic Importance Modeling}  

In this paper, we consider the image's label as its semantic information, such as bird, fox, and human, which corresponds to specific portions of pixels in the image. Generally, a straightforward way to identify semantically critical pixels is through foundation model based segmentation methods \cite{kirillov2023segment}, where all pixels belonging to the object are assigned equal importance. However, this approach is computationally expensive due to the high dimensionality of pixels. In addition,  pixels within the same object may have unequal importance in representing the most critical information about the object \cite{zhou2016learning}. 

To address these issues, we propose an importance modeling framework in the feature domain. By leveraging the spatial correlations in the VAE-encoded latent feature $\bm{y}$, we directly quantify the importance of each feature element to the semantic label. The details are introduced as follows.


\begin{itemize}
	\item \textbf{Spatial property of feature $\bm{y}$.} As shown in Fig. \ref{Neural_Arch2},  when the image is processed through the function $F(\cdot)$, CNN layers maintain the spatial hierarchy of the original input, meaning that the relative positions of objects within the image are preserved in the feature maps. This phenomenon is so-called \emph{spatial consistency} property  \cite{roh2021spatially} of CNNs.  Following this property, we can  model the spatial importance of features within a 2-dimensional space. Without loss of generality, we define an importance matrix $\bm{I}  \in \mathbb{R}^{W_y\times H_y }$ with its element $I_{ij} \in [0,1]$. The importance matrix are same across all the  maps of feature $\bm{y}$.
	
 \begin{figure*}[t]
	\normalsize
	\setlength{\abovecaptionskip}{+0.3cm}
	\setlength{\belowcaptionskip}{-0.1cm}
		\centering
		
		\includegraphics[width=0.7\linewidth]{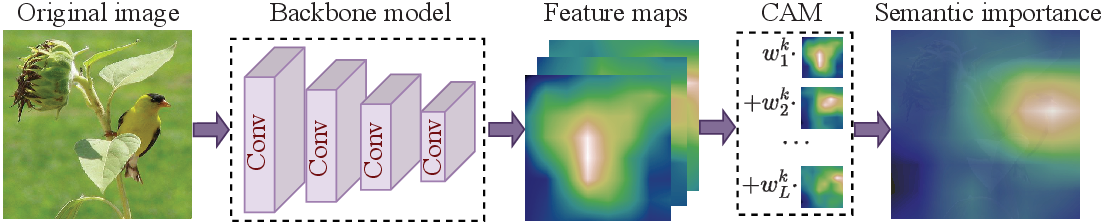}
		\caption{Semantic importance modeling using the CAM. In this example, we blend the importance matrix with the resized image sample  for better illustrations.  The output of the CAM scheme highlights the semantically critical regions w.r.t. object label.}
		\captionsetup{justification=justified}
		\label{CAM}
	\end{figure*}

	\item \textbf{Semantic importance modeling.} We adopt the well-known \emph{class activation mapping} (CAM) \cite{zhou2016learning} to model the spatial importance matrix  $\bm{I}$.  As shown in Fig. \ref{CAM}, the key of CAM is to generate a heatmap that represents the  importance of the spatial grids contributing to the semantic label.  
	First,   a backbone model (e.g. pretrained ResNet-50),  consisting of CNN layers, is utilized to extract the feature maps of images.
	For an image, let $\bm{f}\in \mathbb{R}^{W_f \times H_f \times C_f}$ denote the extracted feature from backbone model  with $W_f$, $H_f$, and $C_f$ being the width, height, and number of feature maps, respectively. Let $\bm{f}_c \in\mathbb{R}^{W_f\times H_f}, c\in[C_f],$ denote the $c$-th feature map of $\bm{f}$.  When performing a  classification task, the extracted  feature $\bm{f}$ will be fed into a fully-connected neural network with softmax function to output the probability of each label.  Specifically, the probability of label $k$, denoted by $P_{k}$, is calculated as 
	\begin{align}
		P_k=\frac{e^{S_k}}{\sum_{k}e^{S_k}},  \text{with}\ S_k= \sum_{c}w_{c}^{k}\sum_{i,j} f_{ijc},
	\end{align}
	where $f_{ijc}$ denotes the $(i,j,c)$-th element of feature $\bm{f}$ and $w_{c}^k$ represents the parameter of the fully-connected neural network for label $k$. Essentially, $w_{c}^k$ indicates the importance of $c$-th feature map  for class $k$.  Hence, the importance matrix of class $k$ is calculated as a weighted  summation of the feature maps, i.e.,
	\begin{align}
		\bm{I}^k = \sum_{c} w_{c}^k \bm{f}_c.
	\end{align}
	Then, each  element of $ \bm{I}^k$ is normalized into the interval $[0,1]$. 
	Finally,  the normalized matrix  $\bm{I}^{k}$ is up-sampled to the spatial size of the feature $\bm{y}$, i.e. $(W_y,H_y)$.
	
	The CAM process outputs the  importance matrices $\{\bm{I}^k\}$ of all the possible object labels. For the image with multiple objects, the transmitter 
	can flexibly choose the importance matrices regarding to the interested objects. 
	In this paper, we take one-label image as an example. Hence, we choose the semantic  importance matrix with the largest detection probability, i.e.,
	\begin{align}
		\bm{I}=\bm{I}^k, \text{~with~} k=\arg \max ~P_k.
 		\end{align}

\end{itemize}

\subsection{Feature Filtering}
With the obtained importance matrix $\bm{I}$, we define an index set as
\begin{align}\label{position}
	\mathbb{P} \triangleq \left\{(i,j,c)|I_{ij} \geq \alpha, \  i \in [W_y], j \in [H_y], c \in[C_y] \right\},
\end{align}
where $ \alpha \geq 0$ denotes the filtering threshold. 
Only the feature elements  belonging to set $\mathbb{P}$, i.e., $y_{ijc}, (i,j,c) \in \mathbb{P}$, are selected for transmission. The filtering process can be expressed as follows,
\begin{align} \label{filtering}
		\tilde{y}_{ijc}= \left\{ \begin{array}{ll}
			y_{ijc}, & \text{if}\ (i,j,c)\in \mathbb{P},\\
			0, & \text{otherwise},
		\end{array}\right.
\end{align}
where $\tilde{y}_{ijc}$ is the $(i,j,c)$-th element of feature $\tilde{\bm{y}}$. 
Then, the filtered feature $\tilde{\bm{y}}$ is quantized into $\hat{\bm{y}}$ and then  encoded into a bit stream for transmission.

\begin{remark}
The proposed filtering process is a controllable procedure aimed at balancing transmission latency and visual quality through the  adjustment of the threshold  $\alpha$. When $\alpha=0$, all visual information is transmitted, rendering the proposed framework equivalent to a data-oriented design. The  selection of $\alpha$ within the range of zero to one results in the loss of certain visual information while effectively reducing transmission latency. Hence, optimizing $\alpha$ is essential to balance visual fidelity and transmission, as will be detailed in Section V.
\end{remark}

\subsection{Side Information Transmission}
To transmit the filtered feature $\tilde{\bm{y}}$, 
the  index set, $\mathbb{P}$, is  treated as a side information and needs to be transmitted to receiver for indicating the positions of the selected features.   To reduce the transmission overhead,  the filtering set $\mathbb{P}$ should be effectively compressed and encoded into bits. According to \eqref{position}, it is observed that only spatial grids $\{(i,j)\}$ of  $\mathbb{P}$   are required for transmission.  Hence, we can construct a binary mask $\bm{m}_{\mathbb{P}}\in \mathbb{R}^{W_y\times H_y \times 1}$, whose $(i,j)$-th image pixel is set to one if $(i,j)\in \mathbb{P}$ or zero otherwise. Then, run-length encoding \cite{zehavi1988runlength} is adopted to effectively encode the mask image into bit stream $\bm{b}_{\mathbb{P}}\in \{0,1\}^{B_{\mathbb{P}}}$.  Note that after  efficient compression, $B_{\mathbb{P}}$ is much smaller than the number of bits for pixels, and can be neglected.  The receiver upsamples the received mask $\bm{m}_{\mathbb{P}}$ to the image size, i.e.,  $\bar{\bm{m}}\in\{0,1\}^{W\times H \times 3}$, to generate images.

\section{Visual Information Fidelity}
 
In this section, we design the GVIF metric to quantify the visual fidelity of the generated images.  We first establish statistical models of image features and their distortions, and then incorporate a well-known HVS model. Based on these models, we derive the GVIF metric. 

\subsection{Statistical Model of Image Features}

Inspired by the  traditional VIF metric, the proposed GVIF metric evaluates image quality through an \emph{ information-theoretic reference-distortion} framework \cite{sheikh2006image,sheikh2005information}. The \emph{reference model} captures the statistical information of original image features, while the \emph{distortion model} simulates degradation processes by quantifying how these features are altered by the Gen-SemCom system.  However, current statistical models for image analysis are limited to the frequency domain \cite{sheikh2006image,sheikh2005information}, leaving a gap in feature-based quality assessment.   To address this issue, we derive the statistical model of image features using the VAE-based hyper prior model introduced in Section II. 


\subsubsection{Reference model}
According to the hyper prior model \cite{Balle2018},  the  feature elements $\{y_{ijc}\}$ for representing  an  image can be modeled as i.n.i.d. conditionally Gaussian variables, i.e., $y_{ijc} | \bm{\theta} \sim \mathcal{N}(0,\theta^2_{ijc})$, where the  standard deviation $\theta_{ijc}$ is the $(i,j,c)$-th element of parameter $\bm{\theta}$.  
Formally,  random variable $y_{ijc}$ can be characterized by a GSM model \cite{sheikh2006image}, which is  the product of  a scalar random variable and a  Gaussian random variable with zero mean and  variance one, i.e.,
\begin{align} \label{model_f}
	y_{ijc}=\theta_{ijc}\cdot u_{ijc}, \forall i \in [W_y],  y\in [H_y], c \in [C_y],
\end{align}
where $\{u_{ijc}\}$ are i.i.d. Gaussian random variables with $u_{ijc}\sim \mathcal{N}(0,1)$, and  are independent of random variables $\{\theta_{ijc}\}$. 
 Note that the variables $\{\theta_{ijc}\}$ in \eqref{model_f} are influenced by the source coder, which inevitably introduces the information loss. However, we aim to obtain the reference model that  simulates the lossless image features for representing the original images. To minimize the distortion from source coder, we select the VAE-based encoder with the  parameter set $\bm{\Phi}^{r}$
w.r.t. a relatively small source distortion, $\mathcal{D}(\bm{\Phi}^r)$. Then, the reference model for the original image is defined as
\begin{align}\label{ref:model}
	y^{r}_{ijc}=\theta_{ijc}^{r}\cdot u_{ijc},
\end{align}
where ${\theta}^{r}_{ijc}=L_2(\hat{\bm{s}};\Omega_2^r)$.  Let $\bm{\theta}^r=\{\theta_{ijc}^r\}$.



		Similar to the image's subbands,  the derived GSM models over feature domain produce heavy-tailed marginal and variance-scaling joint densities, which capture key statistical characteristics of natural images \cite{sheikh2006image,sheikh2005information}.  On one hand, the  key dependency among  image features, such as marginal behaviors, can be modeled using the correlated variables $\{\theta_{ijc}\}$. 
	On the other hand, the  realizations of  random variable $\bm{\theta}$ vary across different image samples, reflecting the distinct contents among images. In conclusions, the realization of random variable $\bm{\theta}$ captures the distinct visual information of the image sample, which will be used to calculate its visual fidelity index in Section IV-B.

%

\subsubsection{Distortion model} 

The purpose of the distortion model is to describe how the image features are disturbed by the VAE, semantic filtering, and generation processes. Here, we ignore the quantization process, since the quantization level is always fixed, while other processes affected by the   NN parameters and the filtering set $\mathbb{P}$ contribute most to the fidelity metric. 

\begin{enumerate}
	\item \textbf{Distortion model of VAE}: 
	According to \eqref{source-rate} and \eqref{model_f},  we observe that all the feature elements extracted from source coders   can be modeled by the GSM model. The source rates of the source coders lie in the scaling of variances $\{\theta^2_{ijc}\}$. Larger $\theta^2_{ijc}$ values correspond to more information retained for image representation.  Hence, we can simply establish the  distortion model from a particular VAE-based source coder  as a scaling model, i.e.,
	\begin{align}\label{compressor}
	 y_{ijc}^{c}=\beta_{ijc}\cdot y^{r}_{ijc},
	\end{align}
	where $0<\beta_{ijc}\leq1$ is a random variable determined by the  source encoder  and image sample.   Denote $\bm{\beta}$ as the collection of scaling variables, i.e.,  $\bm{\beta}=\{\beta_{ijc}\}$.

	\item \textbf{Distortion model of semantic filtering}: Based on the filtering process in \eqref{filtering}, we have 
	\begin{align}\label{filter}
		y_{ijc}^{s}= \left\{ \begin{array}{ll}
			y_{ijc}^c, & \text{if}\ (i,j,c)\in \mathbb{P},\\
			 0, & \text{otherwise}.
		\end{array}\right.
	\end{align}
	\item \textbf{Distortion model of generation process}: For the generative image $\tilde{\bm{x}}$, there exists a  feature  $\bm{y}^{g}$  such that $\tilde{\bm{x}}=F^{-1}(\left\lceil \bm{y}^g\right\rfloor;\Phi_D)$.  According to Section II-D, the diffusion-based process will anchor the feature elements in set $\mathbb{P}$, while generating other feature elements independently  through diffusion-based model. Hence, the generated feature element $y_{ijc}^{g}$ follows
		\begin{align} 
	\label{ge_fe1}	y_{ijc}^{g}&= y_{ijc}^s=
			y_{ijc}^c,\ \text{if}\ (i,j,c)\in \mathbb{P},\\
     \label{ge_fe2}  y_{ijc}^{g} &\text{ is independent of }\ y_{ijc}^c, \ \text{if}\ (i,j,c) \notin \mathbb{P}.
	\end{align}
	\textbf{Validation:} See Appendix A. 
\end{enumerate}

\subsubsection{HVS model}
The model is designed to replicate the way the human brain perceives and processes visual information,  such as object recognition and depth perception. The HVS model can be approximated as a ``Gaussian channel'' with \emph{visual noise} \cite{sheikh2006image,sheikh2005information}, which imposes  limits on how much information can flow through it.

 In this paper, we explore the application of the HVS model in the feature domain to measuring the information of images.  The visual noise can be modeled as a   stationary Gaussian noise with zero mean and  variance ${\gamma}^2$, which is added to each feature element \cite{sheikh2006image,sheikh2005information}. With the HVS model, the distorted feature $y_{ijc}^g$ in \eqref{ge_fe1}-\eqref{ge_fe2}  and the reference feature $y_{ijc}^{r}$ can be respectively disturbed by 
\begin{align}
	\label{HVS:1} g_{ijc}^d&=y_{ijc}^g+n_{ijc},\\
	\label{HVS:2}  g_{ijc}^{r}&=y_{ijc}^{r}+\hat{n}_{ijc},
\end{align}
where the visual noises $n_{ijc}, \hat{n}_{ijc} \sim \mathcal{N}(0,{\gamma}^2)$ are assumed to be independent of $y_{ijc}^{r}$ and $y_{ijc}^g$. Denote $\bm{g}^d=\{g^d_{ijc}\}\in \mathbb{R}^{W_y\times H_y\times C_y}$, $\bm{g}^{r}=\{g^{r}_{ijc}\}\in \mathbb{R}^{W_y\times H_y\times C_y}$, $\bm{n}=\{n_{ijc}\}$, $\hat{\bm{n}}=\{\hat{n}_{ijc}\}$.

\subsection{GVIF Definition, Properties, and Computation}

With the reference, distortion, and HVS models described above, the GVIF metric for the GenAI-enabled system is derived as follows. Similar to the traditional VIF metric\cite{sheikh2005information}, GVIF utilizes  mutual information as the metric   to quantify the amount of information that can be extracted from the output of the HVS. Note that we are interested in the visual fidelity  of a particular  reference-distortion image pair, rather than the average quality of the ensemble of images. Hence, the conditional mutual information given the image sample and distortion processes is utilized to calculate the GVIF. 

 \begin{figure*}[t]
	\normalsize
	\setlength{\abovecaptionskip}{+0.3cm}
	\setlength{\belowcaptionskip}{-0.1cm}
	\centering
	\includegraphics[width=0.8\linewidth]{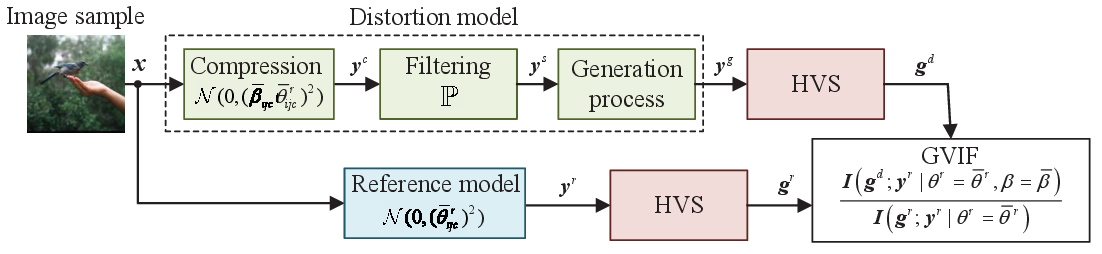}
	\caption{Computation procedures of the GVIF metric.}
	\captionsetup{justification=justified}
	\label{fidelity}
\end{figure*}

\subsubsection{GVIF Metric}  
Let $\bar{\bm{\theta}}^{r}$ and $\bar{\bm{\beta}}$ be the \emph{realizations} of random variables $\bm{\theta}^{r}$ and $\bm{\beta}$, respectively.  Given the statistical parameters $\bar{\bm{\theta}}^{r}$, $\bar{\bm{\beta}}$, and the deterministic set $\mathbb{P}$, we assume the stochastic process governing the distorted feature is ergodic. This enables that mutual information can be calculated using  statistics from a single image sample, bypassing the need for ensemble averages over multiple realizations. 
Then, the conditional mutual information between the distorted feature and its original counterpart is calculated as  $\bm{I}\left(\bm{g}^g;\bm{y}^{r}|\bm{\theta}^{r}=\bar{\bm{\theta}}^{r},\bm{\beta}=\bar{\bm{\beta}}\right)$.   For the reference image sample, the conditional mutual information is calculated as $\bm{I}\left(\bm{g}^{r};\bm{y}^{r}|\bm{\theta}^{r}=\bar{\bm{\theta}}^{r}\right)$. Following \cite{sheikh2005information}, the visual fidelity depends on the ratio of information preserved in the distorted image to that in the reference.   Then, we have the following result.

\begin{proposition}\label{GVIF_P}
For an image sample $\bm{x}$ with the realizations $\bar{\bm{\theta}}^{r}$ and  $\bar{\bm{\beta}}$, and given the variance of visual noise ${\gamma}^2$ and the filtering set $\mathbb{P}$,  the GVIF  can be expressed by
\begin{align}\label{GVIF}
	V(\bar{\bm{\beta}},\mathbb{P};\bm{x}) &\triangleq \frac{\bm{I}\left(\bm{g}^d;\bm{y}^{r}|\bm{\theta}^{r}=\bar{\bm{\theta}}^{r},\bm{\beta}=\bar{\bm{\beta}}\right)}{\bm{I}\left(\bm{g}^{r};\bm{y}^{r}|\bm{\theta}^{r}=\bar{\bm{\theta}}^{r}\right)} \nonumber \\
	&=\frac{\sum_{(i,j,c)\in\mathbb{P}} \log_{2}\left(1+\frac{(\bar{\beta}_{ijc}\bar{\theta}_{ijc}^{r})^2}{{\gamma}^2}\right)}{\sum_{(i,j,c)\in \mathbb{U}} \log_{2}\left(1+\frac{(\bar{\theta}_{ijc}^{r})^2}{{\gamma}^2}\right)},
\end{align}
where $\bar{\theta}_{ijc}^{r}$ and $\bar{\beta}_{ijc}$ are the $(i,j,c)$-th element of tensor parameter $\bar{\bm{\theta}}^{r}$ and $\bar{\bm{\beta}}$, respectively, and $\mathbb{U}=\{(i,j,c)| i \in[W_y], j\in [H_y],c \in [C_y]\}$ is the complete set of dimensions. 
\end{proposition}
\begin{IEEEproof}
	See appendix B. 
\end{IEEEproof}

\begin{figure}[t]
	\centering
		\captionsetup{justification=justified}
	\includegraphics[width=1.0\linewidth]{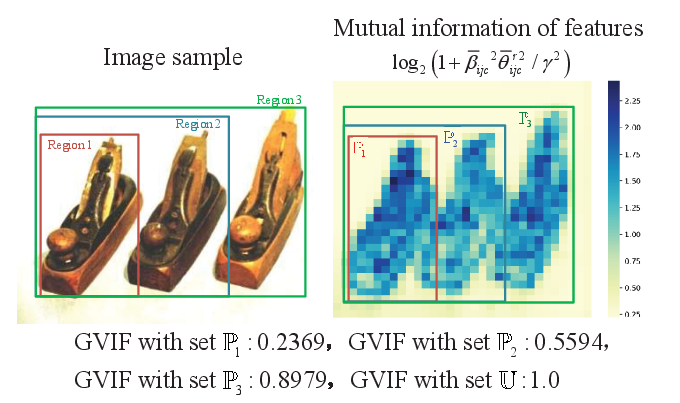}
	\caption{Illustrations of the GVIF's ability for image content understanding. In this image example, we actively choose different set $\mathbb{P}$ to select the image features. For better illustrations, we upsample the sets into image size as regions. We set $\bar{\beta}_{ijc}=1, \forall i,j,c$ and $\gamma=0.1$. The presented mutual information is averaged over feature channels.}
	\captionsetup{justification=justified}
	\label{GVIF_show}
\end{figure}

According to Proposition \ref{GVIF_P}, we have the following observations on the properties of GVIF:
\begin{itemize}
	\item  The GVIF lies in the interval $[0, 1]$ and is a monotonically increasing function w.r.t.  the volume of set $\mathbb{P}$ and $\{\bar{\beta}_{ijc}\}$.  
	First, if the image is undistorted (i.e., $\mathbb{P} = \mathbb{U}$ and $\beta_{ijc} = 1\ \forall (i,j,c)$), the GVIF equals $1$.  
	Second, if no visual information is transmitted (i.e., $\mathbb{P} = \emptyset$), the GVIF equals $0$.  
	Third, if $\mathbb{P} \neq \emptyset$ but critical pixels are completely distorted (i.e., $\beta_{ijc} = 0$), the GVIF also equals $0$.  
	In the second and third cases, the receiver generates images from random noises, resulting in a complete loss of visual fidelity.
	\item  The GVIF demonstrates a strong correlation with the HVS's ability to perceive image's content. Specifically, when the set $\mathbb{P}$ encompasses features with high variance, $(\theta_{ijc}^{r})^2$, indicating presence of critical information such as image object, the GVIF increases quickly.  Conversely, in scenarios where the objects are  missing, the GVIF remains relatively low. To better illustrate this, we present an image sample with different set $\mathbb{P}$, as shown in Fig. \ref{GVIF_show}. It is observed that when the objects are included into the regions for transmission, the GVIF significantly increases. With set $\mathbb{P}_3$, the GVIF is around $0.9$, indicating that the rest of background pixels only contain $0.1$ information of all the  image. 
\end{itemize}


\subsubsection{Computation procedure} Here, we introduce the computation details of the GVIF metric for any reference-distortion  pair. Let $\bm{\Phi}^{r}=\{\Phi_E^r,\Phi_D^r,\Omega_1^c,\Omega_2^r\}$ be the parameter set of the reference source coder with the source distortion being $\mathcal{D}(\bm{\Phi}^r)$. According to the  hyper prior model introduced in Section II, for an image sample $\bm{x}$, the reference parameter $\bar{\bm{\theta}}^r$ is calculated by 
\begin{align}
	\bar{\bm{\theta}}^r= L_2\left(\left \lceil  L_1\left(F(\bm{x};\Phi_E^r);\Omega_1^r\right) \right\rfloor;\Omega^r_2\right).
\end{align} 
For a lossy source coder with  NN parameter set $\bm{\Phi}^{c}=\{\Phi_E^c,\Phi_D^c,\Omega_1^c,\Omega_2^c\}$, we have 
\begin{align}
	\bar{\bm{\theta}}^c= L_2\left(\left \lceil  L_1\left(F(\bm{x};\Phi_E^c);\Omega_1^c\right) \right\rfloor;\Omega^c_2\right),
\end{align}
where $\bar{\bm{\theta}}^c=\{\bar{\theta}^c_{ijc}\}$.
Then, the scaling parameter, $\bar{\bm{\beta}}=\{\bar{\beta}_{ijc}\}$, can be calculated by
\begin{align}
	\bar{\beta}_{ijc}=\frac{\bar{\theta}^{c}_{ijc}}{\bar{\theta}^{r}_{ijc}}.
\end{align}
By substituting the estimated $\bar{\bm{\theta}}^r$ and $\bar{\bm{\beta}}$, and the filtering set $\mathbb{P}$ into \eqref{GVIF}, the GVIF of the distorted image sample  is obtained. 

\section{Application of GVIF to Channel-adaptive Hybrid Gen-SemCom}
In this section, we introduce an application of the proposed GVIF metric, which helps to adapt the Gen-SemCom system  to the channel state.
We formulate the problem based on the proposed GVIF metric and present an efficient algorithm to optimize the system performance accordingly.

\subsection{Problem Formulation}
Our purpose is to adapt the output image quality of the hybrid Gen-SemCom system according to the variations of channel state $h$ and the noise variance $\sigma^2$.  To achieve this, we optimize the set $\mathbb{P}$ and  the scaling variables $\bar{\bm{\beta}}$ to maximize the expected GVIF   for a given receive SNR, $\frac{|h|^2}{\sigma^2}$.  Since the set $\mathbb{P}$ and  the scaling variables $\bar{\bm{\beta}}$ are related to the threshold $\alpha$ in \eqref{position}  and VAE parameter $\bm{\Phi}$,  respectively,  the optimization problem can be formulated as:  
\begin{align}
	\max_{\alpha,\bm{\Phi}}&~~\mathbb{E}_{\bm{x}}\{V(\bm{\Phi},\alpha;\bm{x})\}\label{Problem:1}\\
	\text{s.t.} &~~ \frac{\mathcal{R}(\bm{\Phi},\alpha)}{B\log_{2}(1+|h|^2/\sigma^2)} \leq T_{max}, \tag{\ref{Problem:1}{a}} \label{Problem:11}\\
	&~~\bm{\Phi} \in \{\bm{\Phi}_1,\bm{\Phi}_2,\cdots,\bm{\Phi}_P\}, \tag{\ref{Problem:1}{b}} \label{Problem:12}\\
	&~~ \mathcal{D}(\bm{\Phi})\leq D_0, \tag{\ref{Problem:1}{c}} \label{Problem:13}\\
	&~~ 0 \leq \alpha \leq \alpha_{th}, \tag{\ref{Problem:1}{d}} \label{Problem:14}
\end{align}
where $T_{max}>0$ is the maximal transmission latency, $\alpha_{th}$ represents the maximal threshold for protecting the important semantic information,  and \eqref{Problem:11} represents the average latency constraint with $\mathcal{R}(\bm{\Phi},\alpha) =\mathbb{E}_{\bm{x}}\{B(\bm{\Phi},\alpha;\bm{x})\}$ being the  source rate defined in \eqref{source-rate}.
In constraint \eqref{Problem:12},   the NN parameter $\bm{\Phi}$ is chosen from a well-trained parameter set of $P$ models with different rate-distortion pairs. Constraint \eqref{Problem:13} represents the maximal tolerable distortion on the critical region.

However, problem \eqref{Problem:1} is challenging to solve. Firstly, the expected  GVIF, $\mathbb{E}_{\bm{x}}\{V(\bm{\Phi},\alpha;\bm{x})\}$, and the source rate, $\mathcal{R}(\bm{\Phi},\alpha)$, have no closed-form expressions, due to the intractability of NNs. Secondly, both functions are non-smooth, due to the  set $\mathbb{P}$ w.r.t. the threshold $\alpha$. 
Thirdly, the GVIF is coupled with the filtering threshold $\alpha$ and NN parameters $\bm{\Phi}$, resulting in a non-convex optimization problem. 

\subsection{Solution Algorithm}

To solve problem \eqref{Problem:1}, we propose a two-step algorithm that adaptively optimizes $(\alpha,\bm{\Phi})$  based on the receive SNR, $\frac{|h|^2}{\sigma^2}$.
\begin{enumerate}
	\item \textbf{Fix $\bm{\Phi}$, optimize $\alpha$}: Let $C=B\log_{2}(1+|h|^2/\sigma^2)$. When the NN parameter $\bm{\Phi}$ is fixed, problem \eqref{Problem:1} turns to 
	\begin{align}
		\max_{\alpha}&~~\mathbb{E}_{\bm{x}}\{V(\bm{\Phi},\alpha;\bm{x})\}\label{Problem:2}\\
		\text{s.t.} &~~ \frac{\mathbb{E}_{\bm{x}}\{B(\bm{\Phi},\alpha;\bm{x})\}}{C} \leq T_{max}, \tag{\ref{Problem:2}{a}} \label{Problem:21}\\
		&~~ 0 \leq \alpha \leq \alpha_{th}, \tag{\ref{Problem:2}{b}} \label{Problem:22}
	\end{align}
	By involving the \emph{quadratic penalty function} \cite{boyd2004}, we define a new objective function of problem \eqref{Problem:2} as follows,
	\begin{align}
		\mathcal{L}(\alpha) &\triangleq -\underbrace{\mathbb{E}_{\bm{x}}\{V(\bm{\Phi},\alpha;\bm{x})\}}_{\triangleq \mathcal{G}(\alpha)} \nonumber \\
		&\quad +p \max\{0,\underbrace{\mathbb{E}_{\bm{x}}\{B(\bm{\Phi},\alpha;\bm{x})\}-CT_{max}}_{\triangleq \mathcal{K}(\alpha)}\}^2,
        	\end{align}         
	where $p>0$ is the penalty parameter. Then, 
	 problem \eqref{Problem:2} can be approximated by
	\begin{align} 
				\min_{\alpha}&~~\mathcal{L}(\alpha)\label{Problem:3}\\
		\text{s.t.} 
		&~~ 0 \leq \alpha \leq \alpha_{th} \tag{\ref{Problem:3}{a}} \label{Problem:32}.
	\end{align}
When the penalty parameter, $p$, is large enough, the optimal solution of problem \eqref{Problem:3} approximates to that of problem \eqref{Problem:2}.  According to our numerical experiments,  it is observed that the functions  $\mathbb{E}_{\bm{x}}\{V(\bm{\Phi},\alpha;\bm{x})\}$ and  $\mathbb{E}_{\bm{x}}\{B(\bm{\Phi},\alpha;\bm{x})\}$ are approximately smooth w.r.t.  $\alpha$ as the grid sizes, i.e., $W_y$ and $H_y$ are large enough. Hence, the objective function $\mathcal{L}(\alpha)$ is considered to be continuously differentiable w.r.t. $\alpha$. 
	
	However, the objective function $\mathcal{L}(\alpha)$ and its gradient are still intractable. To overcome this challenge, the zero-order optimization method \cite{liu2020primer} is applied to estimate the gradient using only function evaluations. Then, $\alpha$ can be optimized by applying stochastic gradient descent method. At each iteration $t$, the gradient $\nabla \mathcal{L}(\alpha_t)$ is given by 
	\begin{align}\label{gra:1}
		\nabla \mathcal{L}(\alpha_t)=-\nabla \mathcal{G}(\alpha_t)+2p\max(0,\mathcal{K}(\alpha_t))\nabla \mathcal{K}(\alpha_t).
	\end{align}
The intractable gradients $\nabla \mathcal{G}(\alpha_t)$ and $\nabla \mathcal{K}(\alpha_t)$ in \eqref{gra:1} can be estimated by
	\begin{small}
	\begin{align}
	\label{G_g}	\nabla \mathcal{G}(\alpha_t) &\approx \frac{1}{|\Gamma_t|}\sum_{i=1}^{|\Gamma_t|} \frac{1}{\nu}\left[ V(\bm{\Phi},\alpha_t+\nu m;\bm{x}_i)-V(\bm{\Phi},\alpha_t;\bm{x}_i)\right]m,\\
	\label{K_g}			\nabla \mathcal{K}(\alpha_t) &\approx \frac{1}{|\Gamma_t|}\sum_{i=1}^{|\Gamma_t|} \frac{1}{\nu}\left[ B(\bm{\Phi},\alpha_t+\nu m;\bm{x}_i)-B(\bm{\Phi},\alpha_t;\bm{x}_i)\right]m,
	\end{align}
\end{small}where $\Gamma_t$ denotes the set of mini-batch images at iteration $t$, $\nu>0$ is a smoothing parameter, and $m$ is generated from a uniform distribution centered at zero with radius $1$.  The threshold $\alpha_{t+1}$ is updated by
	\begin{align}\label{Update}
		\alpha_{t+1}=\alpha_t-\eta \nabla \mathcal{L}(\alpha_t),
	\end{align}
	where $\eta>0$ is the step size. To satisfy the constraint in \eqref{Problem:3}, the updated $\alpha_{t+1}$ will be projected into the set $0 \leq \alpha\leq \alpha_{th}$ based on the nearest Euclidean distance.  Finally, we proceed the stochastic gradient descent algorithm until the variables, $\{\alpha_t\}$, converge.

	\item \textbf{Find the best $\bm{\Phi}$}: For ease of analysis, we assume that the source distortions of VAE-based source codings follow $\mathcal{D}(\bm{\Phi}_1)\leq \mathcal{D}(\bm{\Phi}_2)\leq \cdots \leq \mathcal{D}(\bm{\Phi}_P)$. Then, it is easy to find a NN parameter $\bm{\Phi}_g,g\geq 1$, with $\mathcal{D}(\bm{\Phi}_g)\leq D_0$. This reduces the feasible set of $\bm{\Phi}$ to set $\mathbb{G}=\{\bm{\Phi}_1,\bm{\Phi}_2, \cdots, \bm{\Phi}_g\}$.
	With step 1), we  can obtain the solution $\alpha^*(\bm{\Phi}_i)$  for the NN parameter $\bm{\Phi}_i, i=1,2,\cdots,g$. The expected GVIF with $\alpha^*(\bm{\Phi}_i)$ and $\bm{\Phi}_i$, denoted by $\mathbb{E}_{\bm{x}}\{V(\bm{\Phi_i,\alpha^*(\bm{\Phi})};\bm{x})\}$, can be easily  estimated by averaging the GVIF over the image data set. Finally, we find the optimal NN parameter $\bm{\Phi}^*$ with the maximal  expected GVIF, i.e.,
	\begin{align} \label{Optimal}
		\bm{\Phi}^* = \arg \max_{\bm{\Phi}_i\in \mathbb{G}} ~\mathbb{E}_{\bm{x}}\{V(\bm{\Phi,\alpha^*(\bm{\Phi})};\bm{x})\}
	\end{align}

\end{enumerate}
	Based on the above solution, the resultant algorithm for channel-adaptive hybrid Gen-SemCom is summarized in Algorithm 1. The computational complexity of step 1) at each iteration  is calculated as  $O(|\Gamma_t|gW_yH_yC_y)$.
\begin{algorithm}[thb]
	\caption{ Joint optimizations of  source coding and semantic filtering.}
	\label{Algorithm1}
	\hrule
	\vspace{0.3cm}
	\begin{algorithmic}[1]
		\Require NN parameter set $\{\bm{\Phi}_1,\bm{\Phi}_2,\cdots, \bm{\Phi}_P\}$, bandwidth $B$, channel SNR $\frac{|h|^2}{\sigma^2}$, and parameters $\{\gamma,\nu, \eta, T_{max}, D_0\}$.
		\Ensure  $\bm{\Phi}^* $and $ \alpha^*$.
		
		\State Find the feasible set $\mathbb{G}=\{\bm{\Phi}_1,\bm{\Phi}_2, \cdots, \bm{\Phi}_g\}$ with $\mathcal{D}(\bm{\Phi}_g)\leq D_0$. 
		
		\Repeat
		\State Initialize $\alpha_0=0.5$ and fix NN parameter $\bm{\Phi}\in \mathbb{G}$.
		\State Calculate the gradient $\nabla \mathcal{L}(\alpha_t)$ according to \eqref{gra:1}-\eqref{K_g}.
		\State Update the variable  $\alpha_{t+1}$ by \eqref{Update}.
		\Until $||\alpha_{t+1}-\alpha_{t}||<\xi$ and output $\alpha^*(\bm{\Phi})$;
		
		\State For each NN parameter $\bm{\Phi} \in \mathbb{G}$, optimize $\alpha^*(\bm{\Phi})$. 
		\State Estimate the expected GVIFs $\{\mathbb{E}_{\bm{x}}\{V(\bm{\Phi,\alpha^*(\bm{\Phi})};\bm{x})\}\}$.
		
		\State Find the optimal $(\bm{\Phi}^*,\alpha^*(\bm{\Phi}))$ according to \eqref{Optimal}.
	\end{algorithmic}
\end{algorithm}

\section{Experimental Results}

\subsection{Experimental Settings}
\begin{itemize}
	\item \textbf{Model architecture}:  The VAE-based source coder follows the architecture of the hyper prior model \cite{Balle2018}.  We adopt the ResNet-50 model \cite{he2016deep} as the backbone model of the  CAM for spatial importance modelling. We directly adopt the well-trained diffusion model \cite{lugmayr2022repaint,rombach2022high}  with $31$ denosing steps for image generation without any model fine-tuning. 
	\item \textbf{Real dataset}: We test the  hybrid Gen-SemCom system over the well-known ImageNet 2012  dataset \cite{image}, which consists of $50000$  image samples for validation.  All the tested images are resized into $512\times 512 $ pixels. The dataset for training the VAE-based source coders  consists of $100,000$ images sampled from the training dataset of the Open Images Dataset \cite{kuznetsova2020open}.  
	\item \textbf{Parameter settings}: According to the empirical settings in \cite{sheikh2006image}, the variance of HVS noise is set as $\gamma^2=0.1$. The maximal filtering threshold is set as $\alpha_{th}=0.8$. The look-up table of the  source coders consists of $19$ models, with the average PSNR ranging from $27$ dB to $40$ dB.   We utilize the  source coder with the PSNR being $40$ dB as the reference model to extract the reference features. The channel bandwidth is set as $1$ MHz. 
	\item \textbf{Benchmarking schemes}: We consider the classic JPEG2000 \cite{christopoulos2000jpeg2000}  and the VAE-based source coding without filtering as the benchmarking schemes for comparisons. The former is a widely-used source coding scheme in the modern digital transmission system, and the latter has shown the superior compression efficiency compared with traditional schemes, e.g., JPEG and BPG \cite{Bellard2014}.  For better comparisons of generation performance, we also present the images generated only from  the  text prompts, termed as ``Text-to-image Generation", but their transmission latency is  too small and can be neglected.
	
		\item \textbf{Evaluation metrics}:  The proposed GVIF metric is used  to quantitatively track the performance of the Gen-SemCom system. However, it cannot be used for other classic scheme, e.g., JPEG2000. Hence, we still need other metrics as evaluation indicators to demonstrate the performance difference.  Here, we consider two well-known metrics, i.e., mask PSNR  and FID score \cite{zhang2018unreasonable}.  The mask PSNR describes the pixel-level distance of the interested image region, calculated as 
		\begin{small}
		\begin{align}
			\text{mask~PSNR}= 10\log_{10}\left(\frac{255^2}{\frac{1}{\sum\bm{M}_{ijc}}||\bm{M}\odot(\bm{x}-\hat{\bm{x}})||^2}\right),
		\end{align}
	\end{small}where  $\bm{M}\in\{0,1\}^{W\times H \times3}$  is a binary image with value one indicating the interested area and $\bm{M}_{ijc}$ is its $(i,j,c)$-th element.   The binary image $\bm{M}$  can be constructed by upsampling the index set $\mathbb{P}$ w.r.t. $\alpha$, which is described in Section III-C. The FID score is  used to measure the distribution of the generated images compared with the original ones. The lower the FID score, the closer the recovered image is to the original.

\end{itemize}
 \begin{figure*}[t]
	\normalsize
	\setlength{\abovecaptionskip}{+0.3cm}
	\setlength{\belowcaptionskip}{-0.1cm}
	\centering
		\subfigure[]{
		\begin{minipage}[t]{0.33\linewidth}
			\centering
			\includegraphics[width=0.9\linewidth]{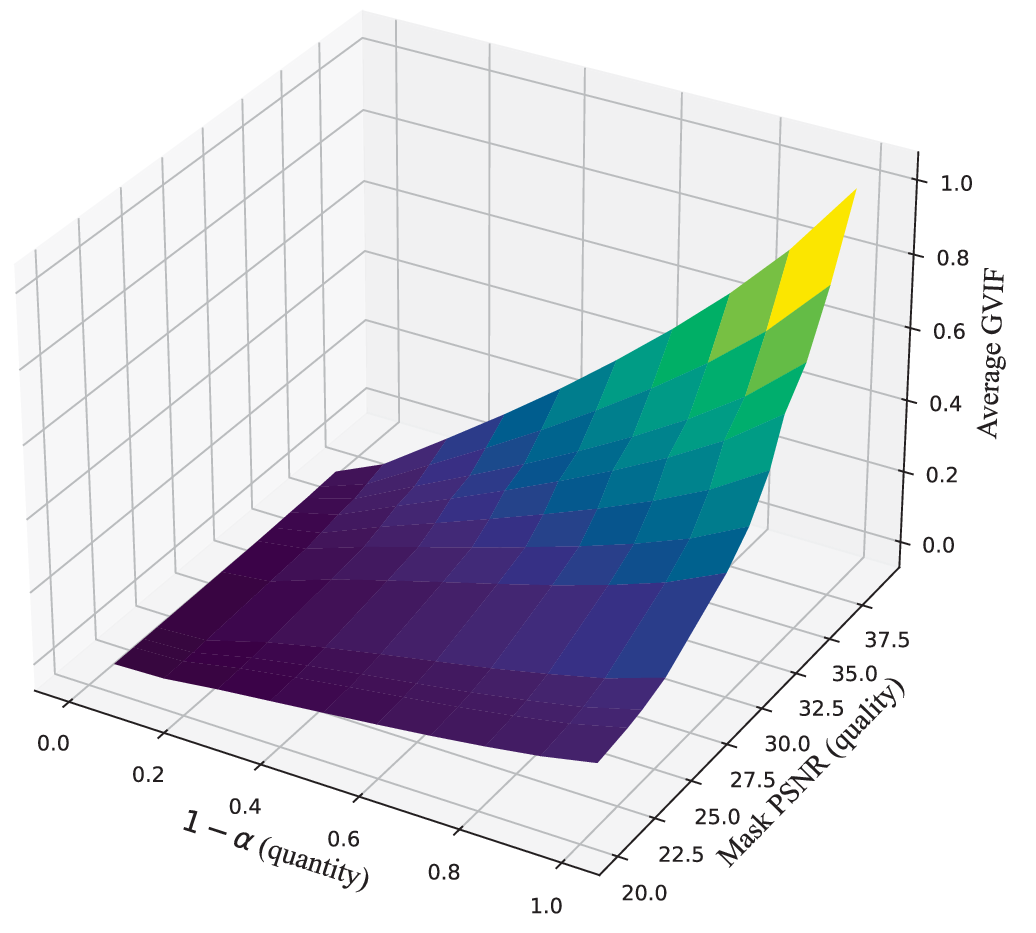}
			
		\end{minipage}%
	}%
	\subfigure[]{
		\begin{minipage}[t]{0.33\linewidth}
			\centering
			\includegraphics[width=0.9\linewidth]{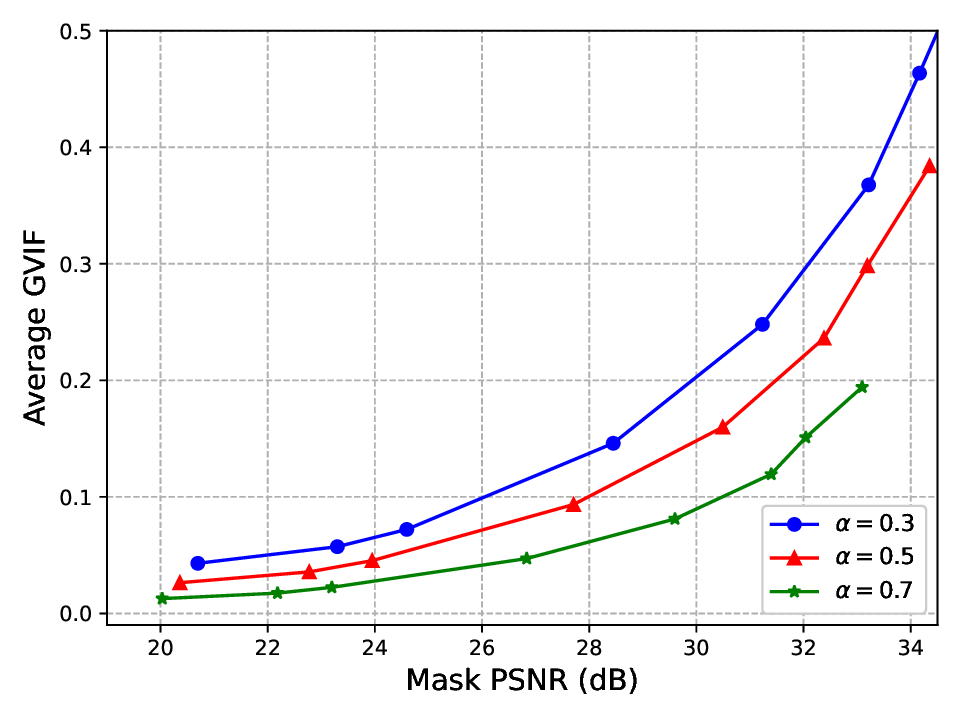}
			
		\end{minipage}%
	}%
	\subfigure[]{
		\begin{minipage}[t]{0.33\linewidth}
			\centering
			\includegraphics[width=0.9\linewidth]{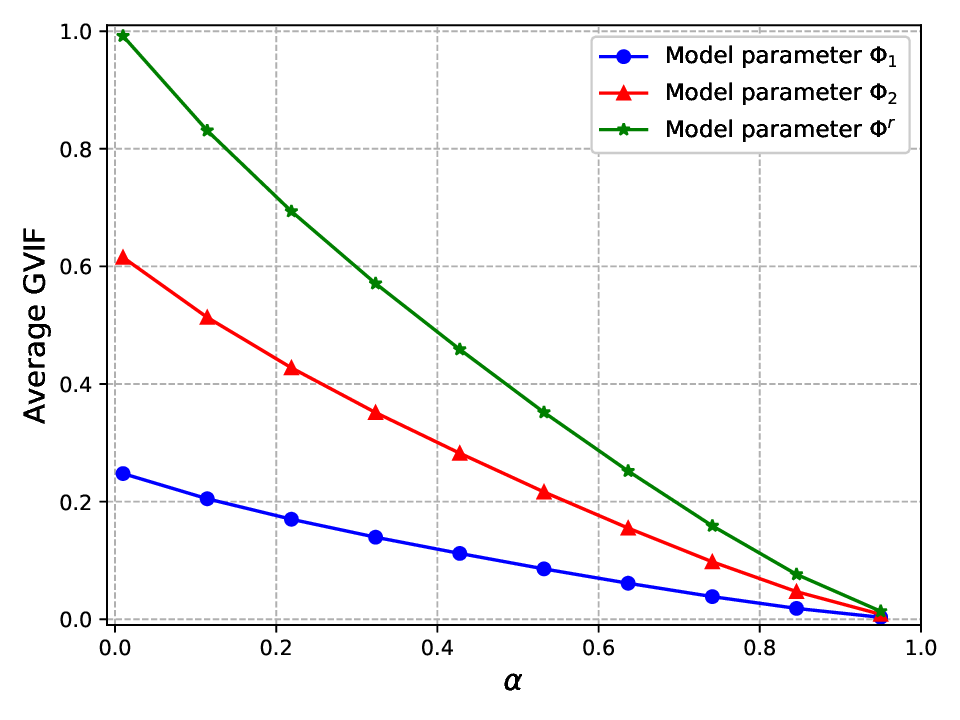}
			
		\end{minipage}%
	}%
	\captionsetup{justification=justified}
	\caption{The  performance of the GVIF metric with different compression distortions and filtering thresholds. }
	\label{property1}
\end{figure*}

 \begin{figure*}[h]
	\normalsize
	\setlength{\abovecaptionskip}{+0.3cm}
	\setlength{\belowcaptionskip}{-0.1cm}
	\centering
	\captionsetup{justification=justified}
	\includegraphics[width=0.9\linewidth]{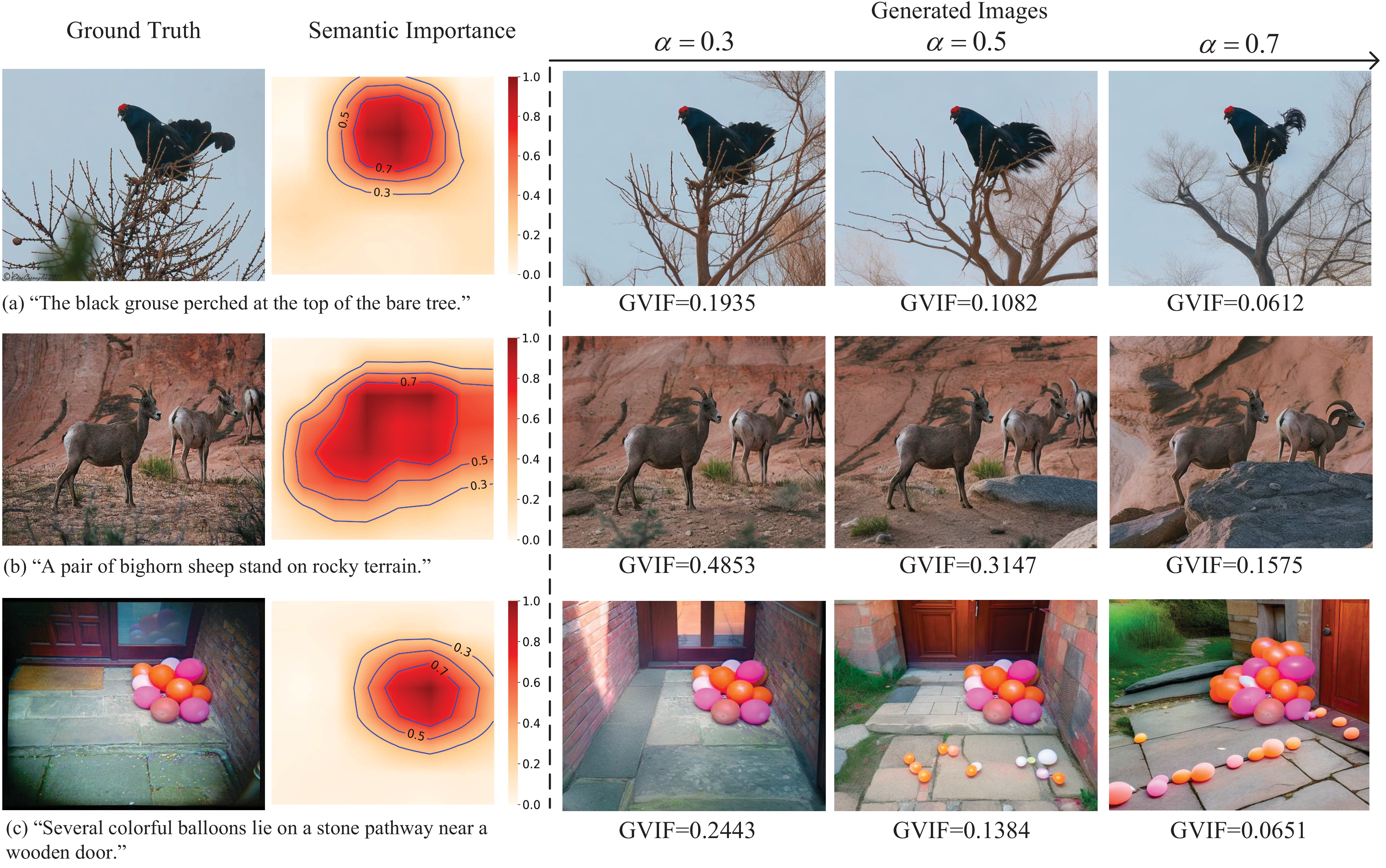}
	\caption{Image examples of the proposed Gen-SemCom system with different  threshold $\alpha$. The mask PSNRs for all the generated images are around $37.5$ dB. The focused semantic information for  images (a)-(c) are ``black grouse",  ``big horn'', and ``balloons'', respectively. }
	\captionsetup{justification=justified}
	\label{Property1}
\end{figure*}

\subsection{Properties of the GVIF metric}

\begin{figure}[t]
	\centering
	\subfigure[]{
		\begin{minipage}[t]{\linewidth}
			\centering
			\includegraphics[width=3.in]{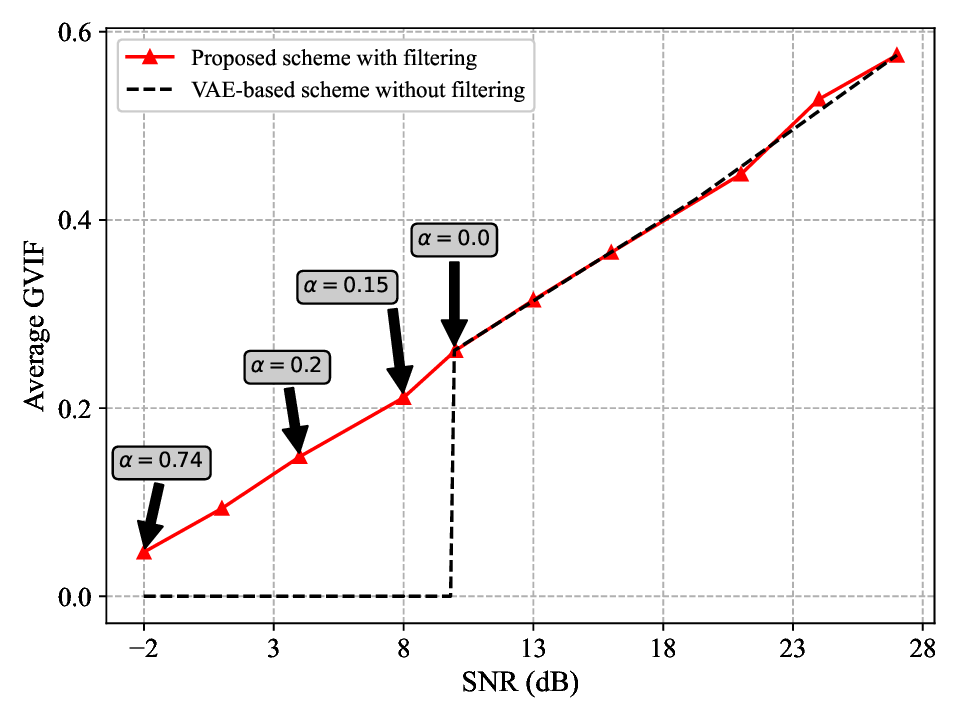}
			
		\end{minipage}%
	}%

	\subfigure[]{
		\begin{minipage}[t]{\linewidth}
			\centering
			\includegraphics[width=3.in]{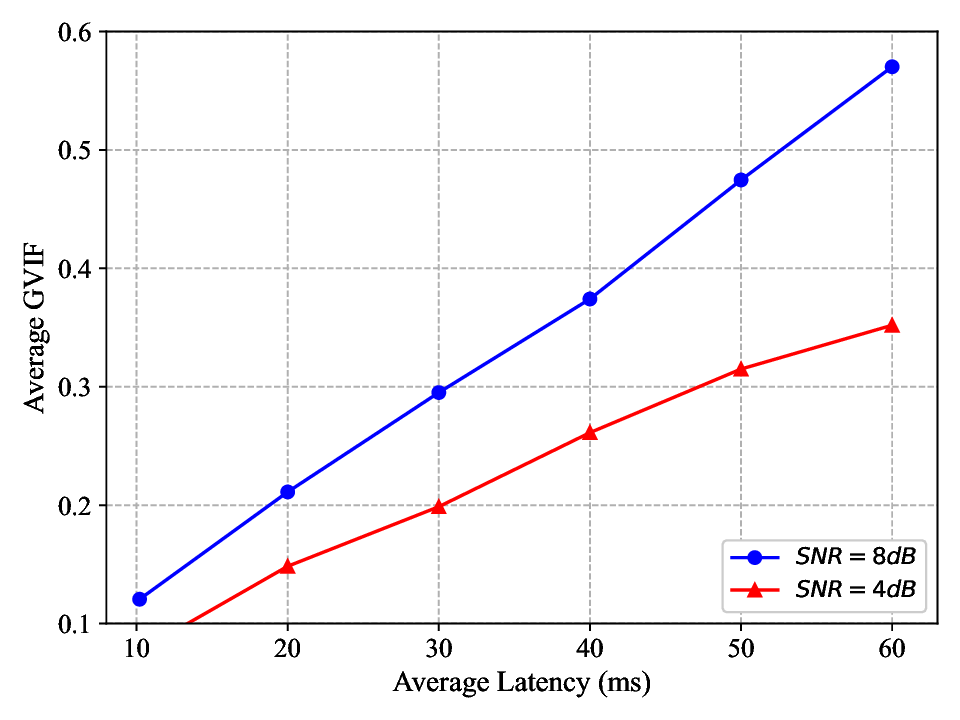}
			
		\end{minipage}%
	}%
	\captionsetup{justification=justified}
	\caption{Performance of the proposed scheme with channel-adaptive design. (a) GVIF v.s. SNR; (b) GVIF v.s. average latency. }
	\label{VIF_p}
\end{figure}

In this subsection, we investigate the properties of the proposed GVIF metric. By changing  the source rates of the  VAE-based source coders, we depict the average GVIF as a function of  the mask PSNR with different thresholds $\alpha$, as shown in Fig. \ref{VIF_p}.  The mask PSNR is related to the source rates of the critical features compressed by VAE-based scheme. The threshold $\alpha$ determines the number of critical features transmitted to the receiver. Both values contribute to the proposed GVIF, which  not only measure the proportion of transmitted image's  content but also quantify their quality distorted by compression process.  This result can be observed in Fig. \ref{VIF_p},  which revevals that the GVIF metric is a monotonously increasing function of both the mask PSNR and the term $1-\alpha$ (which corresponds to the transmitted quantity).  For example, with the mask PSNR being $32$ dB, when  $\alpha$ decreases from $0.7$ to $0.3$, the average GVIF increases around $0.15$ .  An interesting phenomenon is that when $\alpha$ is close to $1$, no matter how advanced the source coder we use, the GVIF remains very low. This observation aligns with the understanding that visual fidelity is primarily determined by the original content. Even though the generated pixels convey the same semantic meaning as the original ones, they can differ significantly in the visual level. Finally, we  conclude that the GVIF metric  comprehensively evaluates the content differences between generated images and original ones, beyond  simple pixel-level comparisons. 

To better illustrate the impact of the semantic filtering process on the SemCom system, we present some image examples in Fig. \ref{Property1}.  It is observed that the larger $\alpha$ is, the more different the generated image is from the original one. For instance, as shown in Fig. \ref{Property1}(c),  the prompt is ``Several pink, orange, and peach balloons lie on a stone pathway near a wooden door.'' Although the generated images capture the accurate semantic information of the original ones, their visual information is decreasing.  This kind of quality loss can not be measured by the commonly-used PSNR metric, because it only focuses on the pixel-level distance. However, our proposed GVIF metric can quantitatively describe the information loss.  In addition, we observe that the semantic importance modeling is able to accurately describe the spatial importance regarding to the semantic information of the image. For instance, in Fig. \ref{Property1}(a), the region with $\alpha=0.3$ almost covers all the features regarding to the black grouse. The increasing of the threshold $\alpha$ degrades the information fidelity, but also decreases the number of the transmitted features. This flexible property of the proposed system facilitates the trade-off between the GVIF and the transmission overhead according to the channel state.

 \begin{figure*}[t]
	\normalsize
	\setlength{\abovecaptionskip}{+0.3cm}
	\setlength{\belowcaptionskip}{-0.1cm}
	\centering
	\subfigure[SNR=-1 dB]{
		\begin{minipage}[t]{\linewidth}
			\centering
			\includegraphics[width=0.9\linewidth]{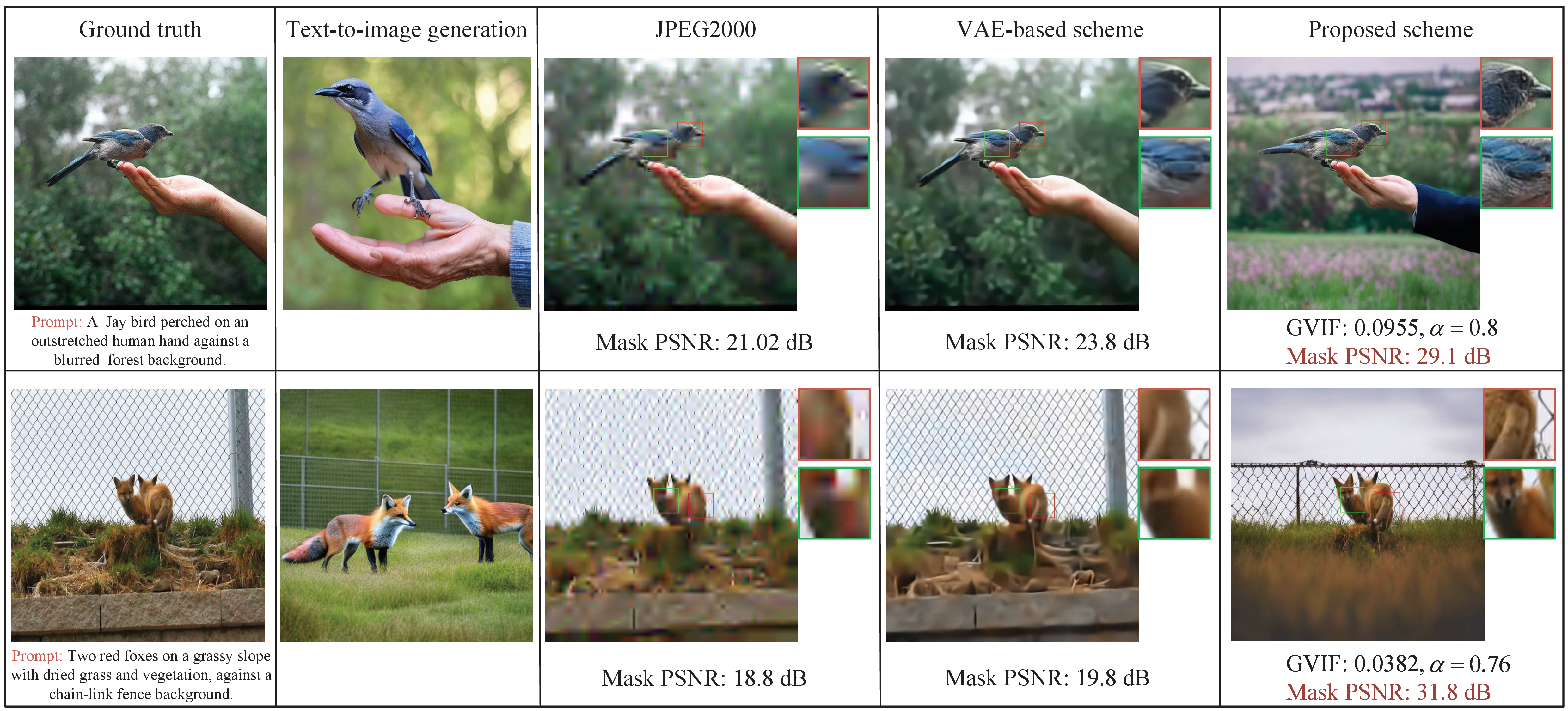}
			
		\end{minipage}%
	}%

	\subfigure[SNR=6 dB]{
		\begin{minipage}[t]{\linewidth}
			\centering
			\includegraphics[width=0.9\linewidth]{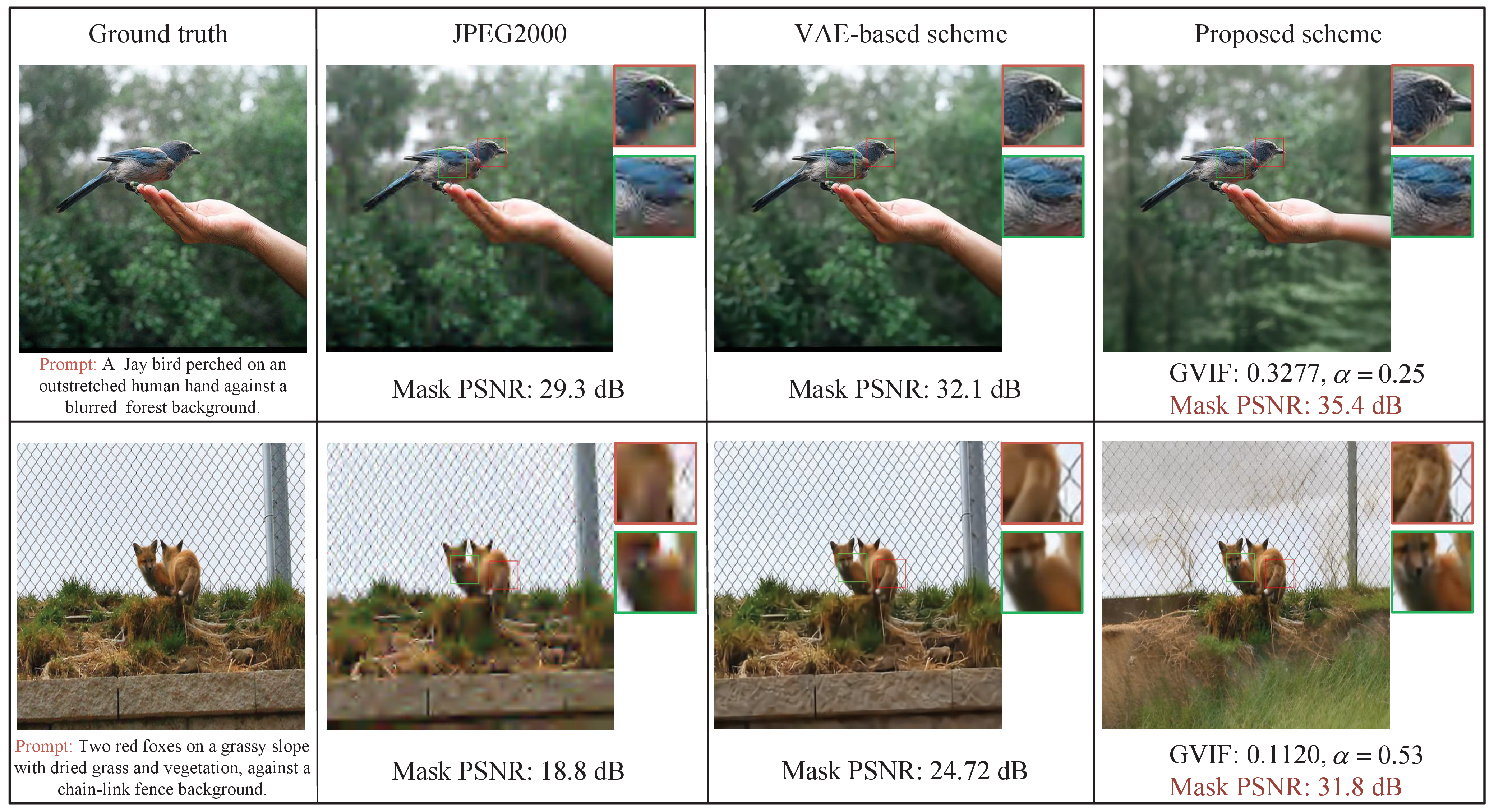}
			
		\end{minipage}%
	}%
	\captionsetup{justification=justified}
	\caption{Image examples of the performance comparisons with different SNRs. }
	\label{visual_1}
\end{figure*}

\subsection{System Performance}
In this subsection, we investigate the performance of the hybrid Gen-SemCom system using the proposed scheme compared with other benchmarking schemes. As shown in Fig. \ref{VIF_p} (a), we plot the average GVIF metric as a function of the channel SNR.  For fair comparisons, we set the minimal mask PSNR of the interested region as $30$ dB.  The maximal transmission latency is set as $20$ ms. From Fig. \ref{VIF_p} (a), we observe that the GVIF of the proposed scheme exhibits a positive correlation with the channel SNR as it increases. A similar phenomenon can be observed in Fig. \ref{VIF_p} (b),  where an increase in transmission latency leads to a rise in the GVIF metric.
An intriguing observation is that 
the  VAE-based source coding scheme without filtering has a cliff-effect phenomenon on the GVIF when the channel SNR is below $10$ dB.  The reason to explain this phenomenon  is that the poor channel conditions are unable to support the transmission of the complete image of high quality.
However, our proposed scheme with filtering process provides the flexibility to regulate the number of the transmitted features, making the GVIF metric more robust to the channel SNR. For instance, when the channel SNR decreases to $-2$ dB, the filtering threshold $\alpha$ increases to $0.74$ to ensure the successful transmission of crucial features related to semantic information. However, it is observed that in the high SNR regions, the proposed scheme  equals to the VAE-based source coding without filtering. 

To better illustrate the performance gain of the proposed scheme, we present some image examples  as shown in Fig. \ref{visual_1}. The average transmission latency for critical pixels is set as $30$ ms.  The prompt of the images in Fig. \ref{visual_1} costs the transmission latency of  around $0.3$ ms, which can be neglected.
Although the ``Text-to-image generation" scheme  can recover  images with minimal transmission latency, the generated images lose almost all  visual fidelity compared to the ground truth. On the other hand, it is observed that our proposed scheme exhibits superior performance, in terms of mask PSNR, compared to the classic JPEG2000 and the VAE-based  scheme. Especially in the low SNR regions, our proposed scheme is shown to protect the key semantic information against the distortions.
 When the channel SNR is $-1$ dB,  an  improvement of around $12$ dB in the mask PSNR can be observed in the ``fox''  image compared to the VAE-based  scheme. 
 When the channel SNR increases to $6$ dB, our proposed scheme can adaptively control the filtering threshold $\alpha$  to maximize the GVIF metric. As shown in Fig. \ref{visual_1}(b), the generated ``bird'' image by our proposed scheme is very close to the ground truth. 

\begin{figure}[t]
	\centering
	\subfigure[]{ 
		\begin{minipage}[t]{\linewidth}
			\centering
			\includegraphics[width=2.6in]{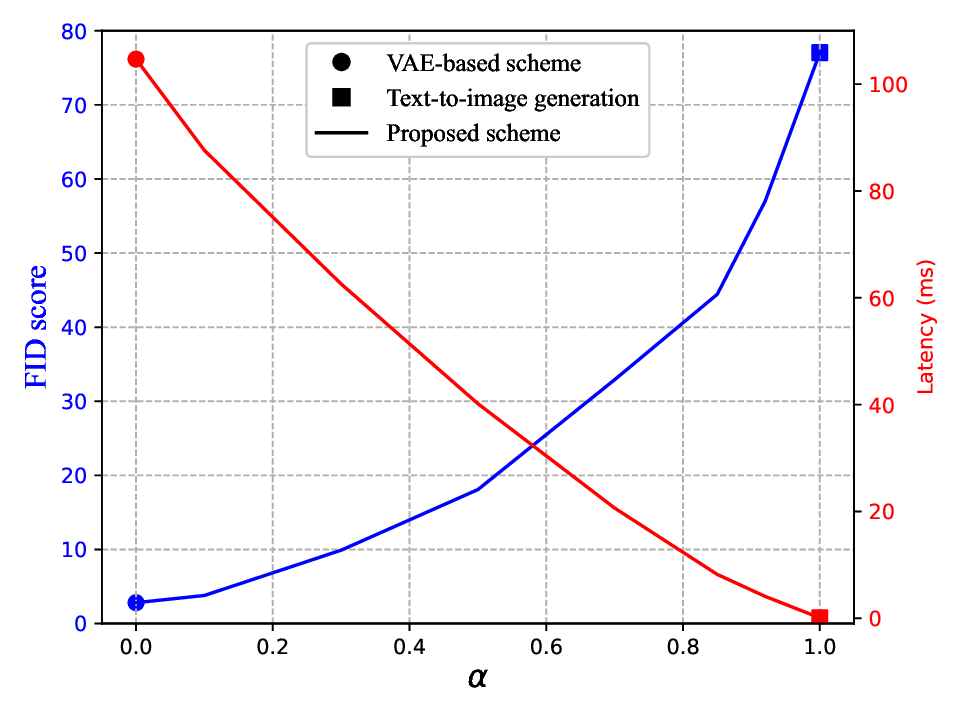}
			
		\end{minipage}%
	}%

	\subfigure[]{
		\begin{minipage}[t]{\linewidth}
			\centering
			\includegraphics[width=2.4in]{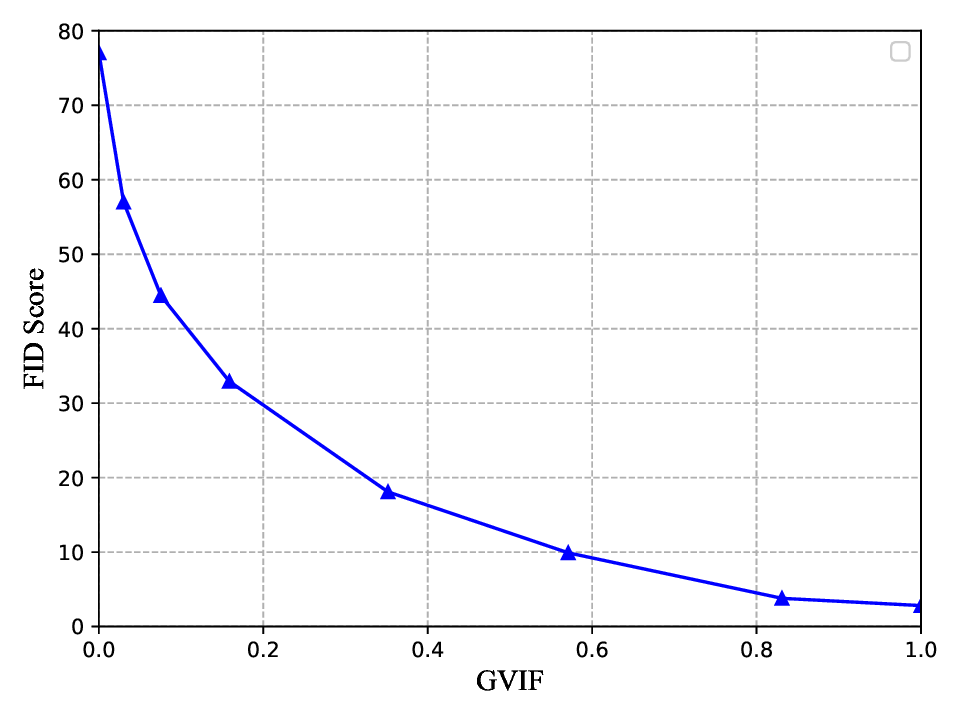}
			
		\end{minipage}%
	}%
	\captionsetup{justification=justified}
	\caption{FID score of the proposed SemCom system. The receive SNR is set as $10$ dB.}
	\label{FID}
\end{figure}

\subsection{FID Score versus GVIF  Metric}

In this subsection, we investigate the FID score of the proposed scheme as shown in Fig. \ref{FID}.   From Fig. \ref{FID}(a), it is observed that the text-to-image generation scheme exhibits the highest FID score, while its transmission latency is the lowest.  The VAE-based source coding scheme exhibits the lowest FID but costs the highest transmission latency. Our proposed scheme provides the flexibility to balance the FID score and the transmission latency by controlling the number of transmitted features.  An interesting phenomenon is that the FID score increases more quickly when $\alpha$ approaches to $1$. For example, when $\alpha$ decreases from $1$ to $0.8$, the FID score decreases by $40$. This phenomenon demonstrates that the distribution of the generated image can closely approximate that of the original image, provided that a few critical features are transmitted.  Next, we  investigate the relationship between the FID score  and the GVIF metric as shown in Fig.\ref{FID} (b). It is observed that the FID score is a monotonously decreasing function of the GVIF, highlighting the capability of GVIF metric to measure images' fidelity.


\section{Concluding Remarks}

In this paper, we proposed a CIE framework for the hybrid Gen-SemCom systems, where only text prompts and critical feature elements are encoded and transmitted to reduce communication overhead. To control the CIE process, we proposed a novel semantic filtering approach that prunes non-critical features based on semantic importance. At the receiver, the prompt and critical features are combined to generate high-quality images using a diffusion-based model.
Furthermore, we proposed the GVIF metric to quantify visual mutual information of the generated images, aligned with human visual perception. The GVIF metric enables joint optimization of encoding and semantic filtering according to  channel state, shifting the focus from traditional rate-distortion tradeoffs to semantic-aware visual quality.
Experimental results validate the GVIF metric’s sensitivity to perceptual quality and demonstrate the  superior performance of the optimized system over benchmarking schemes, achieving higher mask PSNR  and lower FID scores. Future research could  extend the hybrid Gen-SemCom framework to video applications and explore its integration with diverse prompt modalities, such as layout-aware mechanisms, to advance the  fidelity of the generated visual content.

\appendices

\section{Validation of statistical models in \eqref{ge_fe1} and \eqref{ge_fe2}}

In this appendix, we validate the statistic model introduced in \eqref{ge_fe1} and \eqref{ge_fe2} according to the characteristics of diffusion-based model.  
Theoretically, the reverse process of a diffusion model  gradually transforms samples from a Gaussian distribution $\mathcal{N}(0,\bm{I})$, into samples approximating the true conditional distribution $p(\bm{x}|\bm{q},\bar{\bm{m}}\odot \hat{\bm{x}})$. The conditions will anchor the critical pixels and generate rest of pixels. In diffusion models, Gaussian noise is added independently at denoising steps. Hence, the generated image $\tilde{\bm{x}}$ and the true image $\bm{x}$ represent independent samples drawn from the same conditional distribution $p(\bm{x} | \bm{q}, \bar{\bm{m}}_{\mathbb{P}})$.  
  Since the image can be represented by using VAE coder, the generated image $\tilde{\bm{x}}$ can be represented by  a  feature $\bm{y}^{g}$, i.e., $\tilde{\bm{x}}=F^{-1}(\left\lceil \bm{y}^g\right\rfloor;\Phi_D)$. Hence, the generated feature $\bm{y}^{g}$ and the true feature $\bm{y}$ can be viewed as  samples independently drawn from the same distribution $p(\bm{y})$. Since the feature elements  in $\mathbb{P}$ remain unchanged,  the statistical models in \eqref{ge_fe1} and \eqref{ge_fe2} can be proved. 

\begin{figure}[t]
	\centering
	\captionsetup{justification=justified}
	\includegraphics[width=1.0\linewidth]{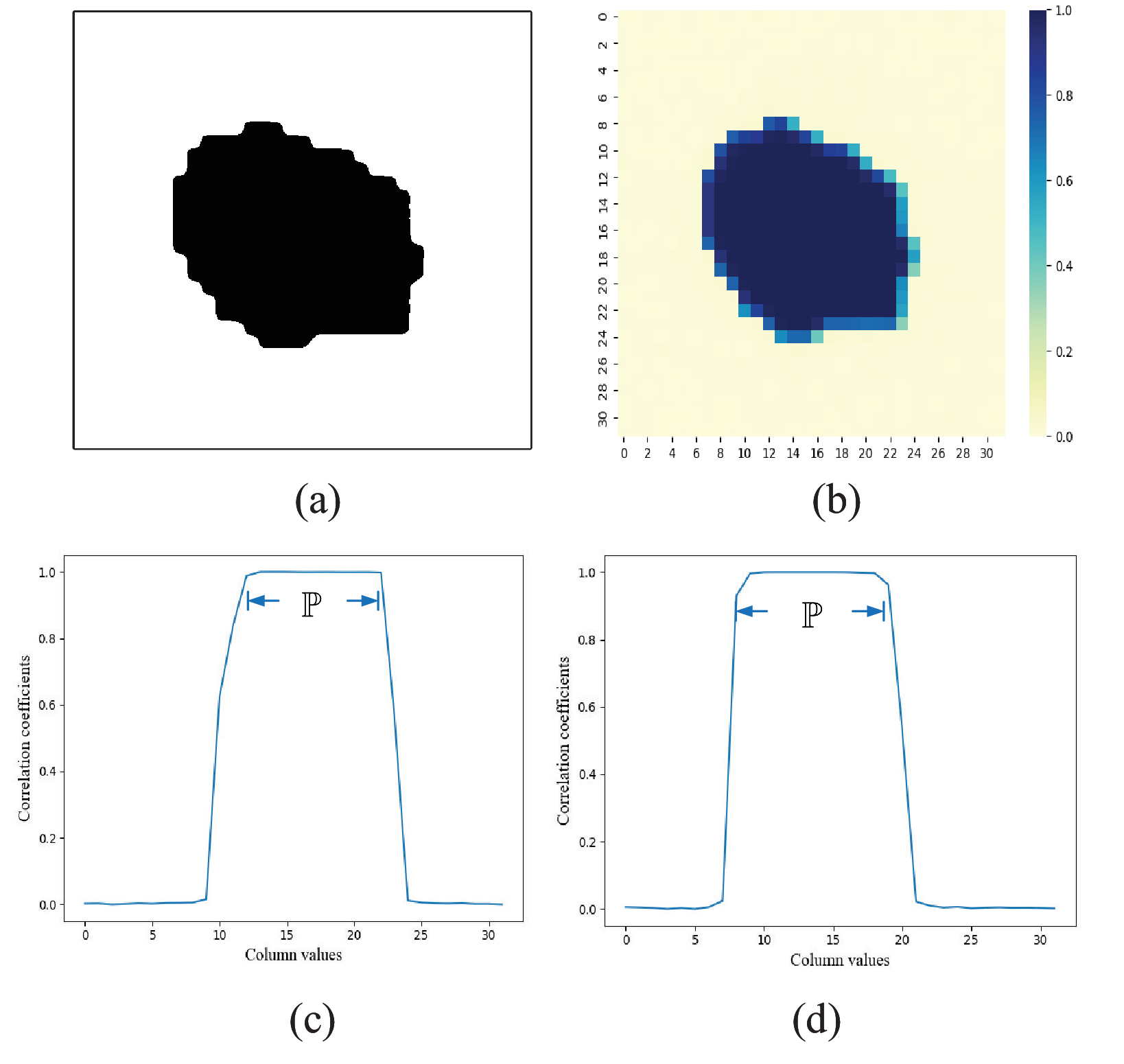}
	\caption{Illustrations of Pearson correlation coefficients over feature maps. (a) Binary mask; (b) Correlation maps; (c) Feature correlations with Row $22$; (d) Feature correlations with Row $12$. }
	\captionsetup{justification=justified}
	\label{statistic1}
\end{figure}

To better validate the statistical model, we conduct some experiments to calculate the correlations of the generated feature $\tilde{\bm{y}}$ and the true feature $\bm{y}$, as shown in Fig. \ref{statistic1}.  We randomly choose $2000$ image samples from ImageNet dataset and calculate the feature samples $\{\bm{y}\}$ by $\bm{y}=F(\bm{x};\Phi_D)$. Each image is fed into the proposed framework with the binary mask shown in Fig. \ref{statistic1} (a) and generate $80$ images $\tilde{\bm{x}}$ by sampling Gaussian noises. The generated feature is obtained by $\tilde{\bm{y}}=F(\tilde{\bm{x}};\Phi_D)$.  Then, we calculate the Pearson correlation coefficients between the generated feature samples $\{\tilde{\bm{y}}\}$ and the true feature samples $\{\bm{y}\}$. The correlation coefficients are averaged over feature maps as shown in Fig. \ref{statistic1} (b)-(d). It is observed that the feature elements within the filtered set $\mathbb{P}$ exhibit high correlation among themselves but appear uncorrelated outside the set. This phenomenon validates the assumption of the statistic model in \eqref{ge_fe1} and \eqref{ge_fe2}. 

\section{Proof of Proposition \ref{GVIF_P}}
According to the HVS model in \eqref{HVS:2}, the conditional mutual information $\bm{I}\left(\bm{g}^{r};\bm{y}^{r}|\bm{\theta}^{r}=\bar{\bm{\theta}}^{r}\right)$ is calculated as 
\begin{align}\label{ref:F}
\bm{I}\left(\bm{g}^{r};\bm{y}^{r}|\bm{\theta}^{r}=\bar{\bm{\theta}}^{r}\right)&\overset{(a)}=\sum_{(i,j,c)\in \mathbb{U}} \bm{I}\left(g_{ijc}^{r};y_{ijc}^{r}|\bm{\theta}^{r}=\bar{\bm{\theta}}^{r} \right), \nonumber \\
&=\frac{1}{2}\sum_{(i,j,c)\in \mathbb{U}}\log_2\left( 1+\frac{(\bar{\theta}_{ijc}^{r})^2}{\gamma^2}\right),
\end{align}
where equality (a) comes from the reference model in  \eqref{model_f} that $\{y_{ijc}^{r}\}$ are independent of each other given  $\bar{\bm{\theta}}^{r}$, and also independent of the visual noises $\{\hat{n}_{ijc}\}$. 

According to the distortion models in \eqref{compressor}-\eqref{ge_fe2} and the HVS model in \eqref{HVS:1}, we have 
\begin{align}
		g_{ijc}^d= \left\{ \begin{array}{ll}
		\beta_{ijc}y_{ijc}^{r}+n_{ijc}, & \text{if}\ (i,j,c)\in \mathbb{P},\\
	y_{ijc}^{g}+n_{ijc}, & \text{otherwise}.
\end{array}\right.
\end{align}
Then, the conditional mutual information, $ \bm{I}\left(\bm{g}^d;\bm{y}^{r}|\bm{\theta}^{r}=\bar{\bm{\theta}}^{r},\bm{\beta}^{r}=\bar{\bm{\beta}}\right)$, is calculated by
\begin{align}\label{Dis:F}
	 &\bm{I}\left(\bm{g}^d;\bm{y}^{r}|\bm{\theta}^{r}=\bar{\bm{\theta}}^{r},\bm{\beta}=\bar{\bm{\beta}}\right) \nonumber \\
	&\quad \quad \quad \quad\overset{(a)}=\sum_{(i,j,c)\in \mathbb{U}} \bm{I}\left(g_{ijc}^{d};y_{ijc}^{r}|\bm{\theta}^{r}=\bar{\bm{\theta}}^{r},\bm{\beta}=\bar{\bm{\beta}} \right), \nonumber \\
	 &\quad  \quad \quad \quad \overset{(b)}=\sum_{(i,j,c)\in \mathbb{P}} \bm{I}\left(g_{ijc}^{d};y_{ijc}^{r}|\bm{\theta}^{r}=\bar{\bm{\theta}}^{r},\bm{\beta}=\bar{\bm{\beta}} \right), \nonumber \\
	 &\quad \quad \quad \quad =\frac{1}{2}\sum_{(i,j,c)\in \mathbb{P}} \log_2\left(1+\frac{(\bar{\beta}_{ijc}\bar{\theta}_{ijc}^r)^2}{\gamma^2} \right),
\end{align}
where equality (a) comes from the fact that $\{y_{ijc}^r\}$ and $\{n_{ijc}\}$ are independent with each other given $\bar{\bm{\theta}}^{r}$, and equality (b) holds because $y_{ijc}^g, \forall (i,j,c) \not \in \mathbb{P}$, are independent of $y_{ijc}^r$, resulting in $\bm{I}(g_{ijc}^d;y_{ijc}^r|\bm{\theta}^r=\bar{\bm{\theta}}^{r},\bm{\beta}=\bar{\bm{\beta}})=0, \forall (i,j,c) \not \in \mathbb{P}$. 
By dividing \eqref{Dis:F} by \eqref{ref:F}, Proposition \ref{GVIF_P} is proved.

\addtolength{\topmargin}{0.02in}	
\IEEEpeerreviewmaketitle
\bibliographystyle{IEEEtran}
\bibliography{semantic}

\begin{thebibliography}{10}
\providecommand{\url}[1]{#1}
\csname url@samestyle\endcsname
\providecommand{\newblock}{\relax}
\providecommand{\bibinfo}[2]{#2}
\providecommand{\BIBentrySTDinterwordspacing}{\spaceskip=0pt\relax}
\providecommand{\BIBentryALTinterwordstretchfactor}{4}
\providecommand{\BIBentryALTinterwordspacing}{\spaceskip=\fontdimen2\font plus
\BIBentryALTinterwordstretchfactor\fontdimen3\font minus
  \fontdimen4\font\relax}
\providecommand{\BIBforeignlanguage}[2]{{%
\expandafter\ifx\csname l@#1\endcsname\relax
\typeout{** WARNING: IEEEtran.bst: No hyphenation pattern has been}%
\typeout{** loaded for the language `#1'. Using the pattern for}%
\typeout{** the default language instead.}%
\else
\language=\csname l@#1\endcsname
\fi
#2}}
\providecommand{\BIBdecl}{\relax}
\BIBdecl

\bibitem{gunduz2022beyond}
D.~G{\"u}nd{\"u}z, Z.~Qin, I.~E. Aguerri, H.~S. Dhillon, Z.~Yang, A.~Yener,
  K.~K. Wong, and C.-B. Chae, ``Beyond transmitting bits: Context, semantics,
  and task-oriented communications,'' \emph{IEEE J. Sel. Areas Commun.},
  vol.~41, no.~1, pp. 5--41, Jan. 2022.

\bibitem{zhang2022toward}
P.~Zhang, W.~Xu, H.~Gao, K.~Niu, X.~Xu, X.~Qin, C.~Yuan, Z.~Qin, H.~Zhao,
  J.~Wei \emph{et~al.}, ``Toward wisdom-evolutionary and primitive-concise
  6{G}: A new paradigm of semantic communication networks,'' \emph{Eng.},
  vol.~8, pp. 60--73, Jan. 2022.

\bibitem{10287247}
Y.~Sun, H.~Chen, X.~Xu, P.~Zhang, and S.~Cui, ``Semantic knowledge base-enabled
  zero-shot multi-level feature transmission optimization,'' \emph{IEEE Trans.
  Wirel. Commun.}, vol.~23, no.~5, pp. 4904--4917, May 2024.

\bibitem{saad2019vision}
W.~Saad, M.~Bennis, and M.~Chen, ``A vision of 6{G} wireless systems:
  Applications, trends, technologies, and open research problems,'' \emph{IEEE
  Netw.}, vol.~34, no.~3, pp. 134--142, June 2019.

\bibitem{zhu2020toward}
G.~Zhu, D.~Liu, Y.~Du, C.~You, J.~Zhang, and K.~Huang, ``Toward an intelligent
  edge: Wireless communication meets machine learning,'' \emph{IEEE Commun.
  Mag.}, vol.~58, no.~1, pp. 19--25, Jan. 2020.

\bibitem{qu2025partialloading}
G.~Qu, Q.~Chen, X.~Chen, K.~Huang, and Y.~Fang, ``{PartialLoading}: User
  scheduling and bandwidth allocation for parameter-sharing edge inference,''
  \emph{arXiv preprint arXiv:2503.22982}, 2025.

\bibitem{10529950}
Z.~Lin, G.~Qu, X.~Chen, and K.~Huang, ``Split learning in 6g edge networks,''
  \emph{IEEE Wirel. Commun.}, vol.~31, no.~4, pp. 170--176, Aug. 2024.

\bibitem{10415235}
Z.~Lin, G.~Zhu, Y.~Deng, X.~Chen, Y.~Gao, K.~Huang, and Y.~Fang, ``Efficient
  parallel split learning over resource-constrained wireless edge networks,''
  \emph{IEEE Trans. Mob. Comput.}, vol.~23, no.~10, pp. 9224--9239, Oct. 2024.

\bibitem{Balle2017}
J.~Ball\'{e}, V.~Laparra, and E.~P. Simoncelli, ``End-to-end optimized image
  compression,'' in \emph{Proc. Int. Conf. Learn. Repres. (ICLR)}, Toulon,
  France, Apr. 2017.

\bibitem{Balle2018}
J.~Ball\'{e}, D.~Minnen, S.~Singh, S.~J. Hwang, and N.~Johnston, ``Variational
  image compression with a scale hyperprior,'' in \emph{Proc. Int. Conf. Learn.
  Repres. (ICLR)}, Vancouver, CA, May 2018.

\bibitem{liu2023learned}
J.~Liu, H.~Sun, and J.~Katto, ``Learned image compression with mixed
  transformer-cnn architectures,'' in \emph{Proc. IEEE/CVF Conf. Comput. Vis.
  Pattern Recognit.}, 2023, pp. 14\,388--14\,397.

\bibitem{10175391}
J.~Huang, D.~Li, C.~Huang, X.~Qin, and W.~Zhang, ``Joint task and data-oriented
  semantic communications: A deep separate source-channel coding scheme,''
  \emph{IEEE Internet Things J.}, vol.~11, no.~2, pp. 2255--2272, Jan. 2024.

\bibitem{dai2022nonlinear}
J.~Dai, S.~Wang, K.~Tan, Z.~Si, X.~Qin, K.~Niu, and P.~Zhang, ``Nonlinear
  transform source-channel coding for semantic communications,'' \emph{IEEE J.
  Sel. Areas Commun.}, vol.~40, no.~8, pp. 2300--2316, June 2022.

\bibitem{li2023fundamental}
D.~Li, J.~Huang, C.~Huang, X.~Qin, H.~Zhang, and P.~Zhang, ``Fundamental
  limitation of semantic communications: Neural estimation for
  rate-distortion,'' \emph{J. Commun. Inf. Net.}, vol.~8, no.~4, pp. 303--318,
  Dec. 2023.

\bibitem{bourtsoulatze2019deep}
E.~Bourtsoulatze, D.~B. Kurka, and D.~G{\"u}nd{\"u}z, ``Deep joint
  source-channel coding for wireless image transmission,'' \emph{IEEE Trans.
  Cog. Commun. Net.}, vol.~5, no.~3, pp. 567--579, May 2019.

\bibitem{10845799}
J.~Huang, K.~Yuan, C.~Huang, and K.~Huang, ``D$^2$-{JSCC}: Digital deep joint
  source-channel coding for semantic communications,'' \emph{IEEE J. Sel. Areas
  Commun.}, vol.~43, no.~4, pp. 1246--1261, Apr. 2025.

\bibitem{erdemir2023generative}
E.~Erdemir, T.-Y. Tung, P.~L. Dragotti, and D.~G{\"u}nd{\"u}z, ``Generative
  joint source-channel coding for semantic image transmission,'' \emph{IEEE J.
  Sel. Areas Commun.}, vol.~41, no.~8, pp. 2645--2657, June 2023.

\bibitem{zhang2018unreasonable}
R.~Zhang, P.~Isola, A.~A. Efros, E.~Shechtman, and O.~Wang, ``The unreasonable
  effectiveness of deep features as a perceptual metric,'' in \emph{Proceedings
  of the IEEE conference on computer vision and pattern recognition}, June
  2018, pp. 586--595.

\bibitem{10531769}
Y.~Zhao, Y.~Yue, S.~Hou, B.~Cheng, and Y.~Huang, ``Lamosc: Large language
  model-driven semantic communication system for visual transmission,''
  \emph{IEEE Trans. Cogn. Commun. Netw.}, vol.~10, no.~6, pp. 2005--2018, Dec.
  2024.

\bibitem{zhang2025semantics}
M.~Zhang, H.~Wu, G.~Zhu, R.~Jin, X.~Chen, and D.~G{\"u}nd{\"u}z,
  ``Semantics-guided diffusion for deep joint source-channel coding in wireless
  image transmission,'' \emph{arXiv preprint arXiv:2501.01138}, 2025.

\bibitem{grassucci2023generative}
E.~Grassucci, S.~Barbarossa, and D.~Comminiello, ``Generative semantic
  communication: Diffusion models beyond bit recovery,'' \emph{arXiv preprint
  arXiv:2306.04321}, 2023.

\bibitem{jiang2024large}
F.~Jiang, S.~Tu, L.~Dong, C.~Pan, J.~Wang, and X.~You, ``Large generative
  model-assisted talking-face semantic communication system,'' \emph{arXiv
  preprint arXiv:2411.03876}, 2024.

\bibitem{jiang2025m4sc}
F.~Jiang, S.~Tu, L.~Dong, K.~Wang, K.~Yang, and C.~Pan, ``M4sc: An mllm-based
  multi-modal, multi-task and multi-user semantic communication system,''
  \emph{arXiv preprint arXiv:2502.16418}, 2025.

\bibitem{10726905}
L.~Dong, F.~Jiang, Y.~Peng, K.~Wang, K.~Yang, C.~Pan, and R.~Schober, ``Lambo:
  Large {AI} model empowered edge intelligence,'' \emph{IEEE Commun. Mag.},
  vol.~63, no.~4, pp. 88--94, 2025.

\bibitem{lin2025hsplitlora}
Z.~Lin, Y.~Zhang, Z.~Chen, Z.~Fang, X.~Chen, P.~Vepakomma, W.~Ni, J.~Luo, and
  Y.~Gao, ``Hsplitlora: A heterogeneous split parameter-efficient fine-tuning
  framework for large language models,'' \emph{arXiv preprint
  arXiv:2505.02795}, 2025.

\bibitem{rombach2022high}
R.~Rombach, A.~Blattmann, D.~Lorenz, P.~Esser, and B.~Ommer, ``High-resolution
  image synthesis with latent diffusion models,'' in \emph{Proc. IEEE/CVF Conf.
  Comput. Vis. Pattern Recognit.}, June 2022, pp. 10\,684--10\,695.

\bibitem{zhao2023survey}
W.~X. Zhao, K.~Zhou, J.~Li, T.~Tang, X.~Wang, Y.~Hou, Y.~Min, B.~Zhang,
  J.~Zhang, Z.~Dong \emph{et~al.}, ``A survey of large language models,''
  \emph{arXiv preprint arXiv:2303.18223}, 2023.

\bibitem{heusel2017gans}
M.~Heusel, H.~Ramsauer, T.~Unterthiner, B.~Nessler, and S.~Hochreiter, ``Gans
  trained by a two time-scale update rule converge to a local nash
  equilibrium,'' \emph{Adv. Neural Inf. Process. Syst.}, vol.~30, pp.
  6626--6637, Dec. 2017.

\bibitem{thorsager2024generative}
M.~Thorsager, I.~Leyva-Mayorga, B.~Soret, and P.~Popovski, ``Generative network
  layer for communication systems with artificial intelligence,'' \emph{IEEE
  Networking Letters}, pp. 82--86, Jan. 2024.

\bibitem{du2024generative}
H.~Du, G.~Liu, D.~Niyato, J.~Zhang, J.~Kang, Z.~Xiong, B.~Ai, and D.~I. Kim,
  ``Generative ai-aided joint training-free secure semantic communications via
  multi-modal prompts,'' in \emph{\emph{Proc. IEEE Int. Conf. Acoust., Speech,
  Signal Process. (ICASSP)}}.\hskip 1em plus 0.5em minus 0.4em\relax IEEE,
  April 2024, pp. 12\,896--12\,900.

\bibitem{tariq2023segment}
S.~Tariq, B.~E. Arfeto, C.~Zhang, and H.~Shin, ``Segment anything meets
  semantic communication,'' \emph{arXiv preprint arXiv:2306.02094}, 2023.

\bibitem{kirillov2023segment}
A.~Kirillov, E.~Mintun, N.~Ravi, H.~Mao, C.~Rolland, L.~Gustafson, T.~Xiao,
  S.~Whitehead, A.~C. Berg, W.-Y. Lo \emph{et~al.}, ``Segment anything,'' in
  \emph{Proc. of the IEEE/CVF Int. Conf. on Computer Vision (ICCV)}, Oct. 2023,
  pp. 4015--4026.

\bibitem{zhou2016learning}
B.~Zhou, A.~Khosla, A.~Lapedriza, A.~Oliva, and A.~Torralba, ``Learning deep
  features for discriminative localization,'' in \emph{Proc. IEEE Conf. Comput.
  Vis. Pattern Recognit.}, 2016, pp. 2921--2929.

\bibitem{sheikh2006image}
H.~R. Sheikh and A.~C. Bovik, ``Image information and visual quality,''
  \emph{IEEE Trans. on Image Processing}, vol.~15, no.~2, pp. 430--444, 2006.

\bibitem{chatgpt2023}
\BIBentryALTinterwordspacing
OpenAI, ``{ChatGPT (Mar 14 version) [Large language model]},'' 2023, accessed:
  Mar. 14, 2023. [Online]. Available: \url{https://chat.openai.com}
\BIBentrySTDinterwordspacing

\bibitem{qu2025mobile}
G.~Qu, Q.~Chen, W.~Wei, Z.~Lin, X.~Chen, and K.~Huang, ``Mobile edge
  intelligence for large language models: A contemporary survey,'' \emph{Early
  Access in IEEE Commun. Surv. Tutor.}, 2025.

\bibitem{li2023blip2}
J.~Li, D.~Li, S.~Savarese, and S.~Hoi, ``{Blip-2: Bootstrapping language-image
  pre-training with frozen image encoders and large language models},'' in
  \emph{Proc. Int. Conf. Mach. Learn. (ICML)}, Honolulu, USA, 2023, pp.
  19\,730--19\,742.

\bibitem{witten1987arithmetic}
I.~H. Witten, R.~M. Neal, and J.~G. Cleary, ``Arithmetic coding for data
  compression,'' \emph{Commun. ACM}, vol.~30, no.~6, pp. 520--540, 1987.

\bibitem{Tse2005}
D.~Tse and P.~Viswanath, \emph{Fundamentals of wireless communication}.\hskip
  1em plus 0.5em minus 0.4em\relax Cambridge University Press, 2005.

\bibitem{lugmayr2022repaint}
A.~Lugmayr, M.~Danelljan, A.~Romero, F.~Yu, R.~Timofte, and L.~Van~Gool,
  ``Repaint: Inpainting using denoising diffusion probabilistic models,'' in
  \emph{Proc. IEEE/CVF Conf. Comput. Vis. Pattern Recognit.}, 2022, pp.
  11\,461--11\,471.

\bibitem{gallager1968information}
R.~G. Gallager, \emph{Information theory and reliable communication}.\hskip 1em
  plus 0.5em minus 0.4em\relax New York, NY, USA: Wiley, 1968.

\bibitem{roh2021spatially}
B.~Roh, W.~Shin, I.~Kim, and S.~Kim, ``Spatially consistent representation
  learning,'' in \emph{Proc. IEEE/CVF Conf. Comput. Vis. Pattern Recognit.},
  June 2021, pp. 1144--1153.

\bibitem{zehavi1988runlength}
E.~Zehavi and J.~K. Wolf, ``On runlength codes,'' \emph{IEEE Trans. on Inf.
  Theory}, vol.~34, no.~1, pp. 45--54, Aug. 1988.

\bibitem{sheikh2005information}
H.~R. Sheikh, A.~C. Bovik, and G.~De~Veciana, ``An information fidelity
  criterion for image quality assessment using natural scene statistics,''
  \emph{IEEE Trans. on Image Processing}, vol.~14, no.~12, pp. 2117--2128, Nov.
  2005.

\bibitem{boyd2004}
S.~Boyd and L.~Vandenberghe, \emph{Convex optimization}.\hskip 1em plus 0.5em
  minus 0.4em\relax UK: Cambridge University Press, 2004.

\bibitem{liu2020primer}
S.~Liu, P.-Y. Chen, B.~Kailkhura, G.~Zhang, A.~O. Hero~III, and P.~K. Varshney,
  ``A primer on zeroth-order optimization in signal processing and machine
  learning: Principals, recent advances, and applications,'' \emph{IEEE Signal
  Process. Mag.}, vol.~37, no.~5, pp. 43--54, Sept. 2020.

\bibitem{he2016deep}
K.~He, X.~Zhang, S.~Ren, and J.~Sun, ``Deep residual learning for image
  recognition,'' in \emph{IEEE Conf. Comput. Vis. Pattern Recog. (CVPR)}, June
  2016, pp. 770--778.

\bibitem{image}
J.~Deng, W.~Dong, R.~Socher, L.-J. Li, L.~Kai, and F.-F. Li, ``Imagenet: A
  large-scale hierarchical image database,'' in \emph{2009 IEEE Conf. Comput.
  Vis. Pattern Recog. (CVPR)}, Miami, FL, USA, June 2009, pp. 248--255.

\bibitem{kuznetsova2020open}
A.~Kuznetsova, H.~Rom, N.~Alldrin, J.~Uijlings, I.~Krasin, J.~Pont-Tuset,
  S.~Kamali, S.~Popov, M.~Malloci, A.~Kolesnikov \emph{et~al.}, ``The open
  images dataset v4: Unified image classification, object detection, and visual
  relationship detection at scale,'' \emph{Inter. J. of Comput. Vis.}, vol.
  128, no.~7, pp. 1956--1981, 2020.

\bibitem{christopoulos2000jpeg2000}
C.~Christopoulos, A.~Skodras, and T.~Ebrahimi, ``The {JPEG}2000 still image
  coding system: an overview,'' \emph{IEEE Trans. Cons. Elec.}, vol.~46, no.~4,
  pp. 1103--1127, Nov. 2000.

\bibitem{Bellard2014}
\BIBentryALTinterwordspacing
F.~Bellard, ``{BPG} image format",'' 2014. [Online]. Available:
  \url{http://bellard.org/bpg/}
\BIBentrySTDinterwordspacing

\end{thebibliography}

\end{document}